%
%
\magnification=1200 
%
%
\vsize=9 true in
\hsize=6.5 true in
\tolerance=10000
\raggedbottom
%
%
\def\loadfive{
  \global\font\fiveit=cmti7 at 5pt
  \gdef\loadfive{\relax}
}

\def\loadseven{
  \global\font\sevenit=cmti7
  \gdef\loadseven{\relax}
}

\def\loadnine{
  \global\font\ninerm=cmr9
  \global\font\ninei=cmmi9
  \global\font\ninesy=cmsy9
  \global\font\ninebf=cmbx9
  \global\font\ninett=cmtt9
  \global\font\nineit=cmti9
  \global\font\ninesl=cmsl9
  \gdef\loadnine{\relax}
}



\def\loadeight{
  \global\font\eightrm=cmr8
  \global\font\eighti=cmmi8
  \global\font\eightsy=cmsy8
  \global\font\eightbf=cmbx8
  \global\font\eightex=cmex8
  \global\font\eighttt=cmtt8
  \global\font\eightit=cmti8
  \global\font\eightsl=cmsl8
  \global\font\eightsc=cmcsc8
  \skewchar\eighti='177			
  \skewchar\eightsy='60			
  \hyphenchar\eighttt=-1			

  \gdef\loadeight{\relax}
}

\def\loadten{
  \global\font\tensc=cmcsc10
  \gdef\loadten{\relax}
}


\def\loadeleven{
  \global\font\elevenrm=cmr10 at 11pt		
  \global\font\eleveni=cmmi10 at 11pt		
  \global\font\elevensy=cmsy10 at 11pt		
  \global\font\elevenex=cmex10 at 11pt		
  \global\font\elevenbf=cmbx10 at 11pt		
  \global\font\elevensl=cmsl10 at 11pt		
  \global\font\eleventt=cmtt10 at 11pt		
  \global\font\elevenit=cmti10 at 11pt		
  \global\font\elevensc=cmcsc10 at 11pt
  \skewchar\eleveni='177			
  \skewchar\elevensy='60			
  \hyphenchar\eleventt=-1			

  \gdef\loadeleven{\relax}
}


\def\loadtwelve{
  \global\font\twelverm=cmr12			
  \global\font\twelvei=cmmi12			
  \global\font\twelvesy=cmsy10 at 12pt		
  \global\font\twelveex=cmex10 at 12pt		
  \global\font\twelvebf=cmbx12			
  \global\font\twelvesl=cmsl12			
  \global\font\twelvett=cmtt12			
  \global\font\twelveit=cmti12			
  \global\font\twelvesc=cmcsc10 at 12pt

  \skewchar\twelvei='177			
  \skewchar\twelvesy='60			
  \hyphenchar\twelvett=-1			
  \gdef\loadtwelve{\relax}
}


\def\loadfourteen{
  \global\font\fourteenrm=cmr10 scaled \magstep2     
  \global\font\fourteeni=cmmi10 scaled \magstep2     
  \global\font\fourteensy=cmsy10 scaled \magstep2    
  \global\font\fourteenex=cmex10 scaled \magstep2		
  \global\font\fourteenbf=cmbx10 scaled \magstep2    
  \global\font\fourteensl=cmsl10 scaled \magstep2    
  \global\font\fourteenit=cmti10 scaled \magstep2    
  \global\font\fourteentt=cmtt10 scaled \magstep2		
  \global\font\fourteensc=cmcsc10 scaled \magstep2

  \skewchar\fourteeni='177                    
  \skewchar\fourteensy='60                    
  \gdef\loadfourteen{\relax}
}


\def\loadsixteen{
  \global\font\sixteenrm=cmr10 scaled \magstep3      
  \global\font\sixteeni=cmmi10 scaled \magstep3      
  \global\font\sixteensy=cmsy10 scaled \magstep3     
  \global\font\sixteenex=cmex10 scaled \magstep3     
  \global\font\sixteenbf=cmbx10 scaled \magstep3     
  \global\font\sixteensl=cmsl10 scaled \magstep3     
  \global\font\sixteentt=cmtt12 scaled \magstep3     
  \global\font\sixteenit=cmti10 scaled \magstep3     
  \global\font\sixteenss=cmss10 scaled \magstep3     
  \global\font\fourteensc=cmcsc10 scaled \magstep3

  \skewchar\sixteeni='177                     
  \skewchar\sixteensy='60                     
  \gdef\loadsixteen{\relax}
}


\def\loadtwenty{
  \global\font\twentyrm=cmr17 at 20pt		
  \global\font\twentyi=cmmi12 at 20pt		
  \global\font\twentysy=cmsy10 at 20pt		
  \global\font\twentyex=cmex10 at 20pt          
  \global\font\twentybf=cmbx12 at 20pt		
  \global\font\twentysl=cmsl12 at 20pt		
  \global\font\twentytt=cmtt12 at 20pt		
  \global\font\twentyss=cmss17 at 20pt		
  \global\font\twentyit=cmti12 at 20pt		

  \skewchar\twentyi='177			
  \skewchar\twentysy='60			
  \hyphenchar\twentytt=-1			
  \gdef\loadtwenty{\relax}
}

\catcode`@=11  

\def\eightpt{%
  \loadeight
  \loadfive
  \loadfive
  \textfont0=\eightrm 
  \textfont1=\eighti 
  \textfont2=\eightsy
  \textfont3=\eightex
  \textfont\itfam=\eightit 
  \textfont\slfam=\eightsl 
  \textfont\ttfam=\eighttt 
  \textfont\bffam=\eightbf 
  \scriptfont0=\fiverm  
  \scriptfont1=\fivei   
  \scriptfont2=\fivesy  
  \scriptfont3=\eightex    
  \scriptfont\itfam=\fiveit
  \scriptfont\bffam=\fivebf 
  \scriptscriptfont0=\fiverm
  \scriptscriptfont1=\fivei 
  \scriptscriptfont2=\fivesy
  \scriptscriptfont3=\eightex 
  \scriptscriptfont\itfam=\fiveit
  \scriptscriptfont\bffam=\fivebf
  \def\rm{\fam\z@\eightrm}%
  \def\mit{\fam\@ne}
  \def\oldstyle{\fam\@ne\eighti}
  \def\cal{\fam\tw@}
  \def\it{\fam\itfam\eightit}%
  \def\sl{\fam\slfam\eightsl}%
  \def\tt{\fam\ttfam\eighttt}%
  \def\bf{\fam\bffam\eightbf}%
  \def\sc{\eightsc}%
  \normalbaselineskip=9pt
  \setbox\strutbox=\hbox{\vrule height6.4pt depth2.6pt width0pt}%
  \normalbaselines%
  \rm%
}%

\def\tenpt{%
  \loadten
  \loadseven
  \loadfive
  \textfont0=\tenrm 
  \textfont1=\teni 
  \textfont2=\tensy
  \textfont3=\tenex
  \textfont\itfam=\tenit 
  \textfont\slfam=\tensl 
  \textfont\ttfam=\tentt 
  \textfont\bffam=\tenbf 
  \scriptfont0=\sevenrm  
  \scriptfont1=\seveni   
  \scriptfont2=\sevensy  
  \scriptfont3=\tenex    
  \scriptfont\itfam=\sevenit
  \scriptfont\bffam=\sevenbf 
  \scriptscriptfont0=\fiverm
  \scriptscriptfont1=\fivei 
  \scriptscriptfont2=\fivesy
  \scriptscriptfont3=\tenex 
  \scriptscriptfont\itfam=\fiveit
  \scriptscriptfont\bffam=\fivebf
  \def\rm{\fam\z@\tenrm}%
  \def\mit{\fam\@ne}
  \def\oldstyle{\fam\@ne\teni}
  \def\cal{\fam\tw@}
  \def\it{\fam\itfam\tenit}%
  \def\sl{\fam\slfam\tensl}%
  \def\tt{\fam\ttfam\tentt}%
  \def\bf{\fam\bffam\tenbf}%
  \def\sc{\tensc}%
  \normalbaselineskip=12pt
  \setbox\strutbox=\hbox{\vrule height8.5pt depth3.5pt width0pt}%
  \normalbaselines%
  \rm%
}%

\def\elevenpt{%
  \loadeleven
  \loadnine
  \loadseven
  \textfont0=\elevenrm 
  \textfont1=\eleveni 
  \textfont2=\elevensy
  \textfont3=\elevenex
  \textfont\itfam=\elevenit 
  \textfont\slfam=\elevensl 
  \textfont\ttfam=\eleventt 
  \textfont\bffam=\elevenbf 
  \scriptfont0=\ninerm  
  \scriptfont1=\ninei   
  \scriptfont2=\ninesy  
  \scriptfont3=\elevenex    
  \scriptfont\itfam=\nineit
  \scriptfont\bffam=\ninebf 
  \scriptscriptfont0=\sevenrm
  \scriptscriptfont1=\seveni 
  \scriptscriptfont2=\sevensy
  \scriptscriptfont3=\elevenex 
  \scriptscriptfont\itfam=\sevenit
  \scriptscriptfont\bffam=\sevenbf
  \def\rm{\fam\z@\elevenrm}%
  \def\mit{\fam\@ne}
  \def\oldstyle{\fam\@ne\eleveni}
  \def\cal{\fam\tw@}
  \def\it{\fam\itfam\elevenit}%
  \def\sl{\fam\slfam\elevensl}%
  \def\tt{\fam\ttfam\eleventt}%
  \def\bf{\fam\bffam\elevenbf}%
  \def\sc{\elevensc}%
  \normalbaselineskip=13pt
  \setbox\strutbox=\hbox{\vrule height9.2pt depth3.8pt width0pt}%
  \normalbaselines%
  \rm%
}%

\def\twelvept{%
  \loadtwelve
  \loadten
  \loadseven
  \textfont0=\twelverm 
  \textfont1=\twelvei 
  \textfont2=\twelvesy
  \textfont3=\twelveex
  \textfont\itfam=\twelveit 
  \textfont\slfam=\twelvesl 
  \textfont\ttfam=\twelvett 
  \textfont\bffam=\twelvebf 
  \scriptfont0=\tenrm  
  \scriptfont1=\teni   
  \scriptfont2=\tensy  
  \scriptfont3=\twelveex    
  \scriptfont\itfam=\tenit
  \scriptfont\bffam=\tenbf 
  \scriptscriptfont0=\sevenrm
  \scriptscriptfont1=\seveni 
  \scriptscriptfont2=\sevensy
  \scriptscriptfont3=\twelveex 
  \scriptscriptfont\itfam=\sevenit
  \scriptscriptfont\bffam=\sevenbf
  \def\rm{\fam\z@\twelverm}%
  \def\mit{\fam\@ne}
  \def\oldstyle{\fam\@ne\twelvei}
  \def\cal{\fam\tw@}
  \def\it{\fam\itfam\twelveit}%
  \def\sl{\fam\slfam\twelvesl}%
  \def\tt{\fam\ttfam\twelvett}%
  \def\bf{\fam\bffam\twelvebf}%
  \def\sc{\twelvesc}%
  \normalbaselineskip=14pt
  \setbox\strutbox=\hbox{\vrule height10.0pt depth4.0pt width0pt}%
  \normalbaselines%
  \rm%
}%

\def\fourteenpt{%
  \loadfourteen
  \loadtwelve
  \loadten
  \textfont0=\fourteenrm 
  \textfont1=\fourteeni 
  \textfont2=\fourteensy
  \textfont3=\fourteenex
  \textfont\itfam=\fourteenit 
  \textfont\slfam=\fourteensl 
  \textfont\ttfam=\fourteentt 
  \textfont\bffam=\fourteenbf 
  \scriptfont0=\twelverm  
  \scriptfont1=\twelvei   
  \scriptfont2=\twelvesy  
  \scriptfont3=\fourteenex    
  \scriptfont\itfam=\twelveit
  \scriptfont\bffam=\twelvebf 
  \scriptscriptfont0=\tenrm
  \scriptscriptfont1=\teni 
  \scriptscriptfont2=\tensy
  \scriptscriptfont3=\fourteenex 
  \scriptscriptfont\itfam=\tenit
  \scriptscriptfont\bffam=\tenbf
  \def\rm{\fam\z@\fourteenrm}%
  \def\mit{\fam\@ne}
  \def\oldstyle{\fam\@ne\fourteeni}
  \def\cal{\fam\tw@}
  \def\it{\fam\itfam\fourteenit}%
  \def\sl{\fam\slfam\fourteensl}%
  \def\tt{\fam\ttfam\fourteentt}%
  \def\bf{\fam\bffam\fourteenbf}%
  \def\sc{\fourteensc}%
  \normalbaselineskip=14pt
  \setbox\strutbox=\hbox{\vrule height10.0pt depth4.0pt width0pt}%
  \normalbaselines%
  \rm%
}%

\def\sixteenpt{%
  \loadsixteen
  \loadfourteen
  \loadtwelve
  \textfont0=\sixteenrm 
  \textfont1=\sixteeni 
  \textfont2=\sixteensy
  \textfont3=\sixteenex
  \textfont\itfam=\sixteenit 
  \textfont\slfam=\sixteensl 
  \textfont\ttfam=\sixteentt 
  \textfont\bffam=\sixteenbf 
  \scriptfont0=\fourteenrm  
  \scriptfont1=\fourteeni   
  \scriptfont2=\fourteensy  
  \scriptfont3=\sixteenex    
  \scriptfont\itfam=\fourteenit
  \scriptfont\bffam=\fourteenbf 
  \scriptscriptfont0=\twelverm
  \scriptscriptfont1=\twelvei 
  \scriptscriptfont2=\twelvesy
  \scriptscriptfont3=\sixteenex 
  \scriptscriptfont\itfam=\twelveit
  \scriptscriptfont\bffam=\twelvebf
  \def\rm{\fam\z@\sixteenrm}%
  \def\mit{\fam\@ne}
  \def\oldstyle{\fam\@ne\sixteeni}
  \def\cal{\fam\tw@}
  \def\it{\fam\itfam\sixteenit}%
  \def\sl{\fam\slfam\sixteensl}%
  \def\tt{\fam\ttfam\sixteentt}%
  \def\bf{\fam\bffam\sixteenbf}%
  \def\sc{\sixteensc}%
  \normalbaselineskip=19pt
  \setbox\strutbox=\hbox{\vrule height13.5pt depth5.5pt width0pt}%
  \normalbaselines%
  \rm%
}%

\def\twentypt{%
  \loadtwenty
  \loadsixteen
  \loadfourteen
  \textfont0=\twentyrm 
  \textfont1=\twentyi 
  \textfont2=\twentysy
  \textfont3=\twentyex
  \textfont\itfam=\twentyit 
  \textfont\slfam=\twentysl 
  \textfont\ttfam=\twentytt 
  \textfont\bffam=\twentybf 
  \scriptfont0=\sixteenrm  
  \scriptfont1=\sixteeni   
  \scriptfont2=\sixteensy  
  \scriptfont3=\twentyex    
  \scriptfont\itfam=\sixteenit
  \scriptfont\bffam=\sixteenbf 
  \scriptscriptfont0=\fourteenrm
  \scriptscriptfont1=\fourteeni 
  \scriptscriptfont2=\fourteensy
  \scriptscriptfont3=\twentyex 
  \scriptscriptfont\itfam=\fourteenit
  \scriptscriptfont\bffam=\fourteenbf
  \def\rm{\fam\z@\twentyrm}%
  \def\mit{\fam\@ne}
  \def\oldstyle{\fam\@ne\twentyi}
  \def\cal{\fam\tw@}
  \def\it{\fam\itfam\twentyit}%
  \def\sl{\fam\slfam\twentysl}%
  \def\tt{\fam\ttfam\twentytt}%
  \def\bf{\fam\bffam\twentybf}%
  \def\sc{\twentysc}%
  \normalbaselineskip=24pt
  \setbox\strutbox=\hbox{\vrule height17.1pt depth6.9pt width0pt}%
  \normalbaselines%
  \rm%
}%

\catcode`@=12  

\font\ninerm=cmr9
\font\nineit=cmti9
%
%
\hyphenation{brems-strah-lung}
%
%
\def\cl{CL 0016+16}
\def\rx{RXJ0018.3+1618}
\def\ie{i.e.}
\def\etal{et al.}
\def\SZ{Sunyaev-Zel'dovich}
\def\ginga{{\tenit Ginga}}
\def\axaf{{\tenit AXAF}}
\def\xmm{{\tenit XMM}}
\def\rosat{{\it ROSAT}}
\def\asca{{\it ASCA}}
\def\cpspamin{counts s$^{-1}$ arcmin$^{-2}$}
\def\kmsMpc{km s$^{-1}$ Mpc$^{-1}$}
\def\arcsec{^{\prime\prime}\mskip-8mu.\mskip1mu}
\def\arcmin{^\prime\mskip-5mu}
\def\slantfrac#1#2{\hbox{$\,^#1\!/_#2$}}
\def\onequarter{\slantfrac{1}{4}}
\def\smallline{\smallskip\noindent}
\def\medline{\medskip\noindent}
\def\spose#1{\hbox to 0pt{#1\hss}}
\def\lsim{\mathrel{\spose{\lower 3pt\hbox{$\sim$}}
        \raise 2.0pt\hbox{$<$}}}
\def\gsim{\mathrel{\spose{\lower 3pt\hbox{$\sim$}}
        \raise 2.0pt\hbox{$>$}}}
%
%
\def\aj{AJ}
\def\araa{ARA\&A}
\def\apj{ApJ}
\def\apjl{ApJ}    
\def\apjs{ApJS}
\def\mnras{MNRAS}
\def\nature{Nature}
\def\pasj{PASJ}
\input epsf
\epsfverbosetrue
\def\epsfsize#1#2{\ifnum#1>\hsize\hsize\else#1\fi}
%
%
\nopagenumbers
\headline={\hfil \folio \hfil}
%
%
%
%
\tenpt
\baselineskip=13pt
%
%
%
\centerline{A Measurement of the Hubble Constant from the X-Ray
Properties}
\centerline{and the Sunyaev-Zel'dovich Effect of \cl}

\vskip 20pt

\centerline{John P.~Hughes\footnote{$^1$}
 {Also Harvard-Smithsonian Center for Astrophysics, 60 Garden Street,
Cambridge, MA 02138}}
\centerline {Department of Physics and Astronomy, Rutgers University,}
\centerline {136 Frelinghuysen Road, Piscataway, NJ 08854-8019}
\centerline {E-mail: jph@physics.rutgers.edu}

\vskip 10pt

\centerline{and}

\vskip 10pt

\centerline{Mark Birkinshaw$^1$}
\centerline{Department of Physics, University of Bristol, Bristol, BS8
1TL, UK}
\centerline {E-mail: mark.birkinshaw@bristol.ac.uk}

\vskip 20pt

\par {{\it Received 17 February 1997; accepted \underbar{~~~} January 1998}}

\vskip 20pt

\centerline {{\bf ABSTRACT}}
\vskip 10pt

A value of the Hubble constant has been determined from a comparison
of the X-ray properties and \SZ\ effect of the distant rich cluster of
galaxies \cl.  The cluster, as imaged by the \rosat\ PSPC, is
significantly elliptical and we present the techniques we have
developed to include this in our analysis. Assuming a smooth,
isothermal gas distribution, we obtain a value $H_0 = 47^{+23}_{-15}\,
\rm km\, s^{-1}\, Mpc^{-1}$, where the errors include systematic and
random uncertainties but are purely observational. Systematic errors
in deprojecting the elliptical surface brightness distribution due to
prolate and oblate geometries as well as arbitrary inclination angles,
introduce an additional fractional error of $\pm17$\% in $H_0$.  At
the redshift of \cl\ ($z=0.5455$) the effect of the cosmological
parameters on the derived $H_0$ value is of order 10\%--20\%; we quote
results for $q_0 = 0.1$.  Combining this result with X-ray/SZ-effect
$H_0$ determinations from seven other clusters and taking account of
systematic uncertainties in our models for the cluster atmosphere, we
find an ensemble value of $H_0 = 42 - 61\,\rm km\, s^{-1}\,
Mpc^{-1}$ with an additional random error of $\pm 16\%$.

\vskip 20pt
\par 
{\it Subject headings:} cosmic microwave background -- cosmology:
observations -- distance scale -- galaxies: clusters: individual (\cl)
-- intergalactic medium

\vfill\eject

\centerline {{\bf 1. INTRODUCTION}}
\par

The \SZ\ (SZ) effect (Sunyaev \& Zel'dovich 1972) offers the promise
of directly measuring the distances to clusters of galaxies, thereby
bypassing the standard cosmic distance ladder. The technique relies on
X-ray observations of a cluster atmosphere and microwave measurements
of the distortion by the cluster of the cosmic microwave background
radiation (CMBR).  In contrast to recent Cepheid-based distance
determinations of galaxies, which have reached the limits of current
instrumentation and still extend only to the Virgo (e.g., Pierce
\etal\ 1994, Freedman \etal\ 1994) or Fornax (Madore \etal\ 1996)
clusters, it is possible to detect (and measure) clusters in both the
X-ray and SZ-effect at cosmologically interesting redshifts of
$\sim$1.  The only other known astrophysical techniques for distance
determination that can work effectively at these distances use Type
Ia supernovae (SNe) as standard candles (e.g., Hamuy \etal\ 1995, Riess,
Press, \& Kirshner 1996, Perlmutter \etal\ 1997) or time delay
measurements of gravitational lenses (Kundi\'c \etal\ 1997; Schechter
\etal\ 1997).  SNIa distances are tied to the Cepheid distance scale,
since the absolute peak magnitudes of, at least, a few nearby SNe need to
be calibrated, while the SZ effect and gravitational lens time delay
techniques are completely independent of all rungs of the cosmic
distance ladder.

\par

The SZ effect is a distortion in the 2.7 K CMBR caused by inverse
Compton scattering of the CMBR photons off electrons in the hot gas of
a cluster of galaxies.  This distortion appears as a decrement in the
microwave background in the Rayleigh-Jeans part of the 
spectrum and an increment on the Wien side.  Even for the richest,
most massive and X-ray luminous galaxy clusters the effect is small,
$\Delta I/I \sim 10^{-4}$, and thus sensitive, low noise microwave
observations are required for the detection of the effect. The value
of such measurements has been known for some time: comparison of the
X-ray properties and the SZ effect for a cluster of galaxies may be
used to measure the cluster's distance, and hence the Hubble constant
(Gunn 1978, Silk \& White 1978, Cavaliere, Danese, \& DeZotti 1979,
Birkinshaw 1979).

\par

\cl\ is one of the most distant clusters for which the SZ effect has
been detected.  Here we present the analysis of X-ray and \SZ\ effect
data from this cluster with a fully consistent modeling of the
cluster's atmosphere under the assumptions that the gas is isothermal,
smooth, and follows an isothermal-$\beta$ law profile (modified for
the evident ellipticity of the cluster).  We present a comprehensive
error analysis flowing down observational uncertainties and
instrumental calibration errors to the derived value of the Hubble
constant.  Rephaeli \& Yankovitch (1997) have pointed out the
importance of considering relativistic corrections to the X-ray
bremsstrahlung emissivity and the CMBR spectral distortion when
interpreting the SZ effect in galaxy clusters.  We have corrected an
error in their results for the bremsstrahlung emissivity (which
derives from an earlier misprint in Gould 1980) that results in a
factor of $\sim$2 reduction in the magnitude of the relativistic
corrections.
At \cl's redshift of 0.5455 (Dressler \&
Gunn 1992), the effect of the cosmological parameters ($\Omega_0$ and
$\Lambda_0$) on the derived Hubble constant can be large: of order
10\%-20\%, depending on the actual numerical values assumed for the
parameters. We present results for a Friedmann Universe
($\Lambda_0=0$) with $\Omega_0 = 0.2$ (equivalently $q_0 = 0.1$) which
appears to be favored by recent measurements of the mean mass density
of the Universe (Carlberg \etal\ 1996).  Correction factors are given
for ``adjusting'' $H_0$ to other popular values of $\Omega_0$ and
$\Lambda_0$.

\par

Over the past few years, the X-ray astronomy satellite \rosat\ has
effected a profound change in our view of galaxy clusters.  Its superb
angular resolution and low background have allowed us to peer more
closely into the morphological structure of clusters, revealing a
wealth of new insights. It is clear now that clusters are complex
objects, showing evidence for substructuring and dynamical activity
from merging, in nearly all cases that have been examined in detail.
Although a boon to astrophysicists studying the evolution of clusters,
this complexity exposes a significant problem for distance determination
using the SZ effect, since the technique requires knowledge of the
three-dimensional distribution of electron density (and temperature)
in the target galaxy cluster.  We have found that new methods of
analysis are necessary in order to understand and quantify the
structural data.  In this and future articles we investigate the
effects of the complex spatial structure of galaxy clusters on
the derivation of the Hubble constant from the SZ effect.

\par

The X-ray image of the cluster \cl, in particular, shows highly
significant evidence for ellipticity with a ratio of major to minor
axis length of $\sim$1.2. Given this observed distribution, it is
immediately obvious that an analysis based on azimuthally-averaged
radial surface brightness profiles would be inadequate.  We employ the
more accurate, but computationally more difficult, approach of model
fitting directly to the image plane.  Once the projected distribution
is modeled, however, the question of how to carry out the deprojection
remains. In our previous work on Abell 665 (Birkinshaw, Hughes, \&
Arnaud 1991, hereafter BHA) and Abell 2218 (Birkinshaw \& Hughes
1994), we assumed that the underlying cluster atmospheres were
spherically symmetric, an approximation which cannot be good for \cl.
In this article we relax the assumption of a fully spherically
symmetric gas distribution and introduce the next level of geometric
complexity, by taking the cluster to be an axisymmetric ellipsoidal
system, with the three-dimensional isodensity contours of the
cluster's gas distribution given by concentric similar ellipsoids of
revolution.  For this surface brightness model the contours of
constant X-ray intensity are similar concentric ellipses, and it is
possible to calculate the effects of varying inclination angles on the
derived value of $H_0$ in closed form.  Since a major uncertainty in
the X-ray/SZ-effect distance determination technique arises from
uncertainties in the intrinsic geometry of the cluster, it is critical
to examine the dependencies of derived distances on the assumptions
made.

\par

In the next section we discuss the X-ray and radio observations
used. In \S 3 we describe the basic method of analysis for both
circular and elliptical surface brightness distributions and describe
the results of fits of these models to the X-ray and SZ effect data.
We present the value of the Hubble constant ($H_0$) in \S4 with a
complete error analysis including both statistical and systematic
errors.  The effects of the unknown three-dimensional geometry of the
cluster are also presented in \S4. Concluding remarks and a summary
are contained in \S5.

\bigskip
\centerline {{\bf 2. OBSERVATIONS}}
\centerline{ {2.1. \it{X-Ray}}}

\par

In order to usefully apply the method outlined above, one
needs accurate measurements of the spatial distributions of density
and temperature in the cluster atmosphere. The currently active
satellite missions \rosat\ (Tr\"umper 1983) and
\asca\ (Tanaka, Inoue \& Holt 1994) both provide greatly enhanced
capabilities relative to previous X-ray satellites for observing
these quantities for all known X-ray--emitting clusters. The various
instruments among these satellites, however, are not all equally
useful for determining both the density and temperature structures.
Here we discuss the merits and limitations of each instrument in this
regard.

\par

The \rosat\ Position Sensitive Proportional Counter (PSPC)
(Pfeffermann \etal\ 1986) has good spatial resolution
($\sim$30$^{\prime\prime}$ half power diameter at 1 keV), very low
internal background, and a large field of view ($\sim$2$^\circ$
diameter) for observations in the soft X-ray band (0.2--2.4
keV). During its lifetime (from launch in June 1990 until it ran out
of detector gas four years later), this instrument revolutionized our
view of the morphologies of galaxy clusters and we use it as the
dataset for investigating the density structure of \cl.

\par

The PSPC data also provide modest spectral resolution ($\Delta E/E
\sim 40\%$ [FWHM] at 1 keV), which most usefully constrains the
equivalent column density of absorbing neutral hydrogen along the line
of sight.  This quantity is essential for converting the observed
rates in the detectors to the ``true'' X-ray flux of the cluster. The
PSPC's ability to measure the temperature of the hot intergalactic
medium in rich galaxy clusters is severely limited by the upper energy
cut-off of the \rosat\ telescope which is far below the ``knee'' in
the bremsstrahlung spectrum for $kT \sim 7$ keV. The gas temperature
is therefore measured using \asca\ data. 

\par

The Gas-Imaging Spectrometer (GIS) (Ohashi et al.~1996) on board \asca\
can produce X-ray images of celestial sources over a broad band
(0.7--10 keV) with modest spectral resolution ($E/\Delta E \sim 8\%$
[FWHM] at 6 keV). Note that there are two nominally identical
detectors referred to as GIS2 and GIS3.  Because of its large field
of view ($\sim$40$^\prime$ diameter) and high efficiency ($>$70\% over
1.5--10 keV), it is the instrument of choice for studying the X-ray
spectra of galaxy clusters. The Solid-state Imaging Spectrometer
(SIS), although it provides impressive spectral resolution, is
considerably less efficient than the GIS at detecting harder X-ray
photons ($E\gsim 4$ keV). This makes the SIS less accurate than the
GIS for temperature determination.  Indeed an earlier analysis of \asca\
data on \cl\ (Yamashita 1994) showed that the uncertainty in the mean
temperature from the SIS data was 6--10 times larger than the
uncertainty from the GIS. (The best fit values were statistically
consistent.)  In our spectral analysis we concentrate on joint
spectral fits of the low energy PSPC data with the higher energy GIS
data.

\par

On the other hand, the limited spatial resolution of the \asca\ X-ray
telescope (half-power diameter of $\sim$3$^\prime$), which is further
broadened by the position resolution of the GIS ($\sim$$0\arcmin.5$ at
6 keV) makes this data considerably less useful than the PSPC data for
determining the detailed internal structure of clusters.  It also
means that measurement of the spatial distribution of temperature
remains impossible for such a distant cluster with current
instrumentation.  

\par\medskip

\centerline{ {\ninerm 2.1.1. \nineit{ROSAT PSPC Imaging}}}
\par

We observed \cl\ with the PSPC in 1992 July for an effective live-time
corrected exposure time of 43,157 s. The observation was carried out
in the standard wobbled mode.  Figure 1 shows a central portion of the
field.  The grayscale presents the number of raw detected events over
the 0.4-2.4 keV band in $7\arcsec5$ square pixels.  The contours show
the same data after background subtraction and exposure correction
(using the standard files supplied as part of the \rosat\ standard
processing) and smoothing with an intensity-dependent smoothing
kernel.  This smoothing employed a Gaussian function with standard
deviation increasing from $7\arcsec5$ in the brightest parts of the
image to 1$^\prime$ in the dimmest.

\par

The cluster appears as the clearly extended source near the center of
Figure~1.  At its peak near the center, \cl\ is a factor of 100 times
brighter than the mean background level.  The intensity of the cluster
is roughly equal to the background level at a radius of
$\sim$3$^\prime$ (1.3 Mpc).  Only a few of the numerous serendipitous
sources that appear in the field have been identified; we mention some
of them below.  In our spatial analysis, and when spectra are
extracted for either source or background regions, we exclude circular
regions of radius $52^{\prime\prime}$ centered on these sources.  For
three relatively bright sources (one of which is the unresolved source
north of the cluster), we use a circular radius of $80^{\prime\prime}$
for the exclusion region.

\par

We verified that the standard background file provided an acceptable
fit to the data using our image fitting software (see below).  We
scaled the background image by a multiplicative factor to fit the data
in an annular region covering 5$^\prime$--10$^\prime$ centered on the
cluster, excluding all sources detected by the standard processing,
but including the small contribution in this region from the X-ray
emission of \cl\ itself, based on the model fits determined using the
nominal (i.e., unit scaled) background.  The multiplicative factor
obtained is 97\% $\pm$ 3\%. Over the region of
interest, i.e., the central 10$^\prime$ of the detector, the variation
in background, as well as the vignetting correction, was small,
$<$5\%.

\par

The PSPC data were boresight corrected using the optical or radio
positions of 5 sources that were coincident with unresolved X-ray
sources. Among these were 3 {\it HST} guide stars at positions (all
positions are quoted in epoch J2000 throughout this paper) 00$^{\rm
h}$17$^{\rm m}$22.1$^{\rm s}$,
16$^\circ$30$^\prime$35$^{\prime\prime}$; 00$^{\rm h}$17$^{\rm
m}$59.1$^{\rm s}$, 16$^\circ$40$^\prime$22$^{\prime\prime}$; and
00$^{\rm h}$19$^{\rm m}$11.9$^{\rm s}$,
16$^\circ$18$^\prime$53$^{\prime\prime}$.  These sources all lie
outside the field of view shown in Figure~1.  We also used QSO 0015+162
at position 00$^{\rm h}$18$^{\rm m}$31.9$^{\rm s}$,
16$^\circ$29$^\prime$26$^{\prime\prime}$, which has a redshift
$z=0.554\pm0.002$ that is very nearly the same as \cl\ (Margon,
Downes, \& Spinrad 1983).  This source is the bright point source
$3\arcmin.3$ north of the cluster in Figure~1; we will be discussing it
in some more detail later.  The unidentified radio source number 15 in
the Moffett \& Birkinshaw (1989) VLA survey of \cl\ (at position
00$^{\rm h}$18$^{\rm m}$31.3$^{\rm s}$, 
16$^\circ$20$^\prime$43$^{\prime\prime}$) also appeared as a PSPC
X-ray source and can be located in Figure~1. The X-ray source positions
were taken from the standard processing and compared to the ``true''
(i.e., optical or radio) source positions.  There was a difference of
$0.28\pm0.18$ s in right ascension and $7.3\pm2.1$ arcseconds in
declination with the X-ray positions being east and north of the true
positions. This corresponds to an error in absolute position
reconstruction of $8\arcsec3$, which is within the usual
range for PSPC data. In the following, we quote boresight-corrected
X-ray positions.

\medskip
\centerline{ {\ninerm 2.1.2. \nineit{ROSAT PSPC Spectroscopy}}}

\par

The light curve (count rate versus time) of the entire PSPC field
shows large count rate excursions (up to factors of two) for short
time periods near the beginning and end of most of the
good-time-intervals provided with the \rosat\ standard processing.
These flares are mostly in the 0.4--0.8 keV band and we attribute them
to solar X-ray fluorescent emission of atmospheric nitrogen and oxygen
from the bright limb of the earth.  We cleaned the data of this
contamination by eliminating the time intervals of high background,
thereby reducing the effective live time of the observation, for
spectral studies only, to 38109.0 s.

The cluster spectrum was extracted from within a circular region of
radius 200$^{\prime\prime}$ centered on the location of the peak
cluster brightness and the background came from a concentric annulus
with inner and outer radii 300$^{\prime\prime}$ and
500$^{\prime\prime}$. All sources (other than the cluster) were
excluded from both source and background regions. We also extracted a
spectrum of the AGN, QSO 0015+162, from a circular region of radius
80$^{\prime\prime}$, employing the same annular background
region. Figure~2 shows the spectrum of each of these two sources.  The
observed PSPC count rate of \cl\ (after background subtraction) is
$(8.57\pm0.17) \times 10^{-2}$ s$^{-1}$, while that of QSO 0015+162 is
$(1.20\pm0.08) \times 10^{-2}$ s$^{-1}$. The background counting rates
associated with these sources are 0.016 s$^{-1}$ and 0.003 s$^{-1}$,
respectively.

\medskip
\centerline{ {\ninerm 2.1.3. \nineit{ASCA GIS Spectroscopy}}}

\par

The GIS observed \cl\ in mid July, 1993, during the Performance
Verification phase of the \asca\ mission, for an effective exposure
time $\sim$35000 s. We extracted the data from the \asca\ archive at
the HEASARC and employed standard screening criteria to the data:
avoidance of South Atlantic Anomaly passages, geomagnetic cosmic ray
cut-off rigidity $>$ 6 GeV/c, and elevation angle between the pointing
direction and limb of the Earth $>$ 10$^\circ$. The light curve of the
entire field of view was examined and no rate anomalies were
noted. Background was taken from the high Galactic latitude blank sky
observations available from the \asca\ Guest Observer Facility with
screening criteria matched to be the same as those used for the \cl\
data.  For both source and background fields, rise time information
(in the form of the RTI, or rise-time invariant, values) was used to
reject charged particle (i.e., non-X-ray) events.

\par

In the GIS image the cluster appeared as a slightly extended source.
However, due to its limited angular resolution, only a few 
serendipitous sources were obvious in the GIS map and, more
importantly, the X-ray source QSO 0015+162 was not cleanly resolved
from the cluster. We extracted the spectrum from within a circular
region of radius 5$^\prime$ which included emission from both the
cluster and the AGN. The position and size of this region were chosen
to avoid a couple of background sources toward the south of the
cluster.  Background came from the same region in detector coordinates
as the source. The GIS spectrum is displayed in Figure 2.  The
background subtracted counting rate of \cl\ (the average of GIS2 and
GIS3) was $(4.64\pm 0.13) \times 10^{-2}$ s$^{-1}$.  The background
rate was 0.013 s$^{-1}$.

\medskip
\centerline{ {2.2. \it{OVRO \SZ\ Effect Data}}}
\par
The \SZ\ effect for \cl\ was measured using the 40-m telescope of the
Owens Valley Radio Observatory (OVRO) at 20.3~GHz over the period
1983-1990 (Birkinshaw \etal\ 1997). The method used was similar to
that used to measure the SZ effects of Abell~665 and~2218. The
dual-beam system on the 40-m telescope provides two 
1.78-arcmin FWHM beams separated by 7.15~arcmin in azimuth. Data were
taken using beam-switching and position-switching at seven locations
on a north-south line through the center of \cl\ (as determined from
{\it Einstein Observatory\/} data). 
The measured SZ effect signals therefore represent the difference between the
brightness of the microwave background radiation towards the scan
locations and a weighed average of points in ``reference arcs''
centered 7.15~arcmin from them. These reference arcs are irregularly
sampled because of the varying rate with which the parallactic angle
of the reference beams changes with time, because of the exclusion of
some parts of the reference arcs because of radio source contamination,
and because of the varying weather conditions over the
observations. This irregular sampling causes the measured SZ effects
to be complicated functions of the intrinsic SZ effect of \cl.
\par
Figure~3 shows the measurements at these seven locations in \cl, after
averaging the best data taken over the entire observing period and
correcting for the presence of contaminating radio sources. The
largest radio-source corrections were needed at the points 4 and
7~arcminutes south of the center of the cluster: the errors on the
measurements at these points are large because of the uncertainty in
the corrections. Fortunately, these heavily-contaminated points lie
sufficiently far from the center of \cl\ that the source corrections
have little effect on the fitted amplitude of the cluster SZ effect.
\par

In addition to the statistical errors, the error bars shown in Fig.~3
contain contributions from the year-to-year inconsistency in the data,
uncertainties in the radio source corrections, and errors in the
telescope pointing. These are significant for the points 7~and
4~arcmin south of the cluster center (near a bright radio source), but
make only small corrections to the errors elsewhere. A further
systematic error represented in Fig.~3 is the possible offset in the
zero level of the data, such as might be caused by differential
spillover effects (see Birkinshaw \etal\ 1997). This zero level offset
is best measured by the amount that extreme points in the scan are
offset from the nominal zero level, after correction for the expected
SZ effects at the outer points. For \cl\ the best-fit zero level
offset is $+70 \pm 43 \ \rm \mu K$, reflecting the tendancy of the
outer points in the scan to lie at positive brightness temperatures.
The range of this zero level error is represented by the horizontal
lines in Fig.~3.
\par
A final important source of uncertainty is that arising from the
brightness temperature scale. This was based on an internal noise
source in the receiver, which was independently calibrated using hot
and cold loads, planets, and measurements of unresolved radio
sources. An absolute error of 6\% arises from uncertainty in the beam
response of the telescope, while the calibrated value of the
equivalent brightness of the internal load has a 5\% error. Combining
these errors together, the overall systematic uncertainty in the
brightness temperature scale in Fig.~3 may be as much as 8\%.
\par
It can be seen that \cl\ shows a strong central SZ effect, and that
this SZ effect is extended by more than the 1.78-arcmin FWHM beam of
the 40-m telescope. The measured SZ effect near the center of the
cluster is $-490 \pm 80 \ \rm \mu K$ (with no zero level offset),
about half the intrinsic SZ effect of the cluster.
\bigskip
\centerline {{\bf 3. ANALYSIS}}
\centerline {3.1. {\it{Basic Method}}}

BHA derived the following generalized expressions for the X-ray
surface brightness and the \SZ\ effect from a cluster of galaxies

$$\eqalign{b_X(\theta,\phi) &= {1 \over 4 \pi (1+z)^3} \Lambda_{e0}
      n_{e0}^2 D_A  \int d\zeta f^2_n f_\Lambda \cr
    & \equiv N_X \Theta_X \cr} \eqno(1) $$

$$\eqalign{\Delta T_{RJ}(\theta,\phi) &= -2T_r{k_B T_{e0} \over m_ec^2}
     \sigma_T n_{e0} D_A  \int d\zeta f_n f_T \cr
   &\equiv N_{RJ} \Theta_{RJ}\ , \cr} \eqno(2)$$

\noindent
where $\theta$ and $\phi$ form a Cartesian coordinate system.  In
these equations $T_r$ ($= 2.728$ K, Fixsen \etal\ 1996) is the
temperature of the microwave background radiation, $z$ ($= 0.5455$) is
the cluster redshift, $\Lambda_{e0}$ is the spectral emissivity of the
cluster gas at $T_{e0}$ calculated over the emitted energy range
appropriate for the observed energy range of 0.4--2.4 keV, 
$n_{e0}$ is the electron
number density at the center of the cluster, $T_{e0}$ is the central
electron temperature, $D_A$ is the angular diameter distance of the
cluster, $\sigma_T$ is the Thomson scattering cross-section, $k_B$ is
the Boltzmann constant, $m_e$ is the electron mass, and $c$ is the
speed of light.  The variation of the electron density, temperature,
and the spectral emissivity with (three-dimensional) position in the
cluster is contained in the dimensionless form factors $f_n$, $f_T$,
and $f_\Lambda$.  $\zeta$ is an angular measure of distance along the
line of sight. The structural information on the cluster is contained
in the angles $\Theta_X$ and $\Theta_{RJ}$ and the normalizations
of the X-ray and SZ effects are in $N_X$ and $N_{RJ}$.

\par

Our previous work (BHA; Birkinshaw \& Hughes 1994) considered only
spherical cluster atmospheres.  Here we extend the analysis to
include elliptical models.  For convenience we first present a summary
of the analysis using circular isothermal-$\beta$ models to describe
the gas distribution before proceeding to the more general cases.

\par\medskip
\centerline{\ninerm {3.1.1. {\nineit {Circular Isothermal-$\beta$ Model}}}}

Under the assumption of an isothermal atmosphere ($f_T = 1$,
$f_\Lambda = 1$), with a density distribution given by $n_e = n_{e0} [
1+ (r/r_c)^2]^{-3\beta/2}$ (Cavaliere \& Fusco-Femiano 1976, 1978) we
derive

$$\Theta_X(\theta,\phi) = \sqrt{\pi}\ {\Gamma(3\beta-{1\over2})
  \over\Gamma(3\beta)}\ \theta_C\biggl(1 + {\theta^2 + \phi^2 \over
  \theta_C^2}\biggr)^{{1\over2}-3\beta} \eqno(3)$$

$$\Theta_{RJ}(\theta,\phi) = \sqrt{\pi}\ {\Gamma({3\beta\over2}-{1\over2})
  \over\Gamma({3\beta\over2})}\ \theta_C\biggl(1 + {\theta^2 + \phi^2 \over
  \theta_C^2}\biggr)^{{1\over2}-{3\beta\over2}} \eqno(4)$$

\par\noindent
for the forms of the X-ray and SZ surface brightnesses of the
cluster. The angular diameter distance of the cluster is given by 

$$D_A = \biggl({N_{RJ}^2\over N_X}\biggr) 
  \biggl({m_ec^2 \over k_B T_{e0}}\biggr)^2 
  {\Lambda_{e0} \over 16 \pi T_r^2 \sigma_T^2 (1+z)^3}. \eqno(5)$$

\par
In practice, we analyze the PSPC X-ray image to determine best-fit
values and errors for the quantities $\beta$, $\theta_C$, and $N_X$.
These values are used to construct model SZ profiles (including the
appropriate subtraction of residual cluster emission in the reference
arcs) which are then fit to the OVRO scan data to obtain $N_{RJ}$.  We
make the assumption that the cluster atmosphere is isothermal, \i.e.,
the form factors $f_T$ and $f_\Lambda$ are everywhere equal to
unity. Thus the remaining observable quantity, $k_B T_{e0}$, can be
obtained from \asca\ X-ray spectroscopy.

\par\medskip
\centerline{\ninerm {3.1.2. {\nineit {Elliptical Isothermal-$\beta$ Model}}}}

In this case, the surface brightness of the cluster is assumed to be a
function of $[\theta^2 + (e\phi)^2]/\theta_C^2$, where $\theta$ lies
along the major axis of the cluster, $\phi$ along the minor axis, and
$e$ is the ratio of major to minor axes.  We assume that the three
dimensional structure of the cluster is given by an ellipsoid of
revolution with the symmetry axis lying at inclination angle $i$ to
the line of sight (see Fabricant, Rybicki, \& Gorenstein 1984 for
details on the projection of spheroids). The results then depend on
whether we assume an oblate (with the symmetry axis lying along
$\phi$) or prolate (symmetry axis along $\theta$) geometry. Again an
isothermal atmosphere is assumed.

\par
For an oblate geometry we find

$$\Theta_X(\theta,\phi) = \sqrt{\pi}\ {\Gamma(3\beta-{1\over2})
  \over\Gamma(3\beta)}\ \theta_C {\sqrt{1-e^2 \cos^2i} \over \sin i} 
  \biggl(1 + {\theta^2 + (e\phi)^2
  \over\theta_C^2}\biggr)^{{1\over2}-3\beta}  \eqno(6)$$

$$\Theta_{RJ}(\theta,\phi) = \sqrt{\pi}\ {\Gamma({3\beta\over2}-{1\over2})
  \over\Gamma({3\beta\over2})}\ \theta_C {\sqrt{1-e^2 \cos^2i} \over
  \sin i} \biggl(1 + {\theta^2 + (e\phi)^2 \over
  \theta_C^2}\biggr)^{{1\over2}-{3\beta\over2}}  \eqno(7)$$

while a prolate geometry yields

$$\Theta_X(\theta,\phi) = \sqrt{\pi}\ {\Gamma(3\beta-{1\over2})
  \over\Gamma(3\beta)}\ \theta_C {\sqrt{e^2 - \cos^2i} \over e^2\sin i}
  \biggl(1 + {\theta^2 + (e\phi)^2 \over
  \theta_C^2}\biggr)^{{1\over2}-3\beta}  \eqno(8)$$ 

$$\Theta_{RJ}(\theta,\phi) = \sqrt{\pi}\ {\Gamma({3\beta\over2}-{1\over2})
  \over\Gamma({3\beta\over2})}\ \theta_C {\sqrt{e^2 - \cos^2i} \over
  e^2\sin i}\biggl(1 + {\theta^2 + (e\phi)^2 \over
  \theta_C^2}\biggr)^{{1\over2}-{3\beta\over2}}. \eqno(9)$$ 

\noindent

The equation for $D_A$ is the same as above (eqn. 5).  Here the X-ray
image provides values and errors for the quantities $\beta$,
$\theta_C$, $e$, the position angle of the major axis, and $N_X$ while
fits to the SZ data yield $N_{RJ}$.  We present results for different
assumed inclination angles and for the oblate and prolate geometries.

\par\medskip
\centerline{3.2. {\it {Fits to the X-ray image}}}

The purpose of our image analysis is to explore fits of the preceding
classes of structural models to the \cl\ X-ray data in order to
produce an accurate representation of the density structure of the
cluster.  In addition to determining best-fit values and errors for
the relevant parameters we also would like to determine the
goodness-of-fit of these models.  The standard method of fitting an
azimuthally averaged radial surface brightness profile, as has been
applied most widely to cluster data in the past, is grossly
inadequate to explore the complex structures that now, thanks to
\rosat, are clearly evident in the image of \cl\ and other
clusters. The approach we take here is one that we introduced earlier
(BHA, Birkinshaw \& Hughes 1994) and is based on performing spatial
model fits directly to the two-dimensional image data.  This approach
provides us with the flexibility to explore the complex structures
necessary to understand fully the morphologies of galaxy clusters.

\par

X-ray images of clusters are sparsely filled with most image pixels
containing zero or one detected event and \cl\ is no exception (the
mean number of events per pixel of the image shown in Figure 1 is
$\sim$1). The statistical error associated with counting experiments
of this kind follow a Poisson distribution.  Only in the case of a
large enough number of detected events (typically $\gsim$10) does a
Gaussian distribution serve as an adequate approximation to the
Poisson distribution. Since the usual figure-of-merit function, the
$\chi^2$ statistic, requires that measurement errors be normally
distributed, we were led to derive a different maximum likelihood
estimator for the Poisson distributed error case.

\par

The Poisson probability that $D_{ij}$ events were observed in a given
image pixel, $(i,j)$, for a predicted number of model X-ray events
$M_{ij}$  is $P_{ij} = M_{ij}^{D_{ij}}\,e^{-M_{ij}}/D_{ij}!$.
The likelihood function $L$ is the product of the individual
probabilities $P_{ij}$ over all pixels in the region of the image
being fitted. For reasons having to do with determining confidence
intervals (as discussed below), we choose to minimize the function
$S\equiv -2\ln L$ which is equivalent to maximizing $L$.  Thus our
figure-of-merit function becomes
$$S = \sum_{ij} 2M_{ij} - 2D_{ij}\ln M_{ij}, \eqno(10)$$
where we have dropped all terms that do not depend on the model being
fitted.   We employ the robust downhill simplex method (Press \etal\
1986) for function minimization.
\par

This new estimator, however, does not provide an analytic
goodness-of-fit criterion.  In our work here, we take the following
approach.  Radial profiles about the cluster center of the X-ray data
and best-fit model in four azimuthal sectors (defined by the ordinal
compass directions) are compared in a $\chi^2$ test and only radial bins
containing more than 10 events are used in the calculation.  In the
results presented below, we give the $\chi^2$ values computed in this
manner for the various structural models.

\par

On the other hand, our function $S$ does provide a method for
determining the relative goodness-of-fit between different models and
for the estimation of confidence intervals, through the ``likelihood
ratio test.''  This test is carried out
by comparing the value $S_{\rm min}$, determined by minimizing the
figure-of-merit function over all relevant parameters, with the value
$S_r$, which is determined from a fit where $r$ parameters have values
that differ from the best fit ones.  It can be shown that the
distribution function of $S_r - S_{\rm min}$ tends to a $\chi^2$
distribution with $r$ degrees of freedom (see, for example, Kendall \&
Stuart 1979).

\par

As mentioned earlier, circular regions centered on all non-cluster
X-ray sources were excluded from the fits; these image pixels were
not used in the computation of the likelihood function.  Aside
from this, the fit included all pixels within a $5^\prime$ radius of
the peak cluster emission. All results are quoted below for the same
fixed region. We did explore the effect of fitting the data over a
smaller region in radius ($3^\prime$) and found that the change in
derived parameter values was negligible.  Our model included exposure
correction and convolution (in the Fourier domain) with the on-axis
PSPC point response function (Hasinger \etal\ 1992) calculated for a
photon energy of 1 keV.  This is important, but its effect is not
overwhelming: the FWHM of the \rosat\ PSPC point response function is
$\sim$25$^{\prime\prime}$, which is about \onequarter\ of the FWHM of
the cluster itself.

\par

Fits were carried out first to the circular isothermal-$\beta$ model.
Numerical values of the best-fit parameters are given in Table 1.  The
position of the center of the X-ray cluster is consistent (agrees to
within $<$10$^{\prime\prime}$) with the position of the central bright
galaxy in the cluster (Dressler \& Gunn 1992). The best-fit $\theta_C$
value corresponds to a reasonable physical size of $\sim$340 kpc at
the cluster (for $H_0 = 50$ km s$^{-1}$ Mpc$^{-1}$, $q_0 =
0.1$). However, this fit is not particularly good.  The $\chi^2$ (for
the data and model binned in quadrants) of 140.1 for 91 degrees of
freedom can be formally rejected at the 99.93\% confidence level.  The
error intervals presented in the last column of the table are based on
the likelihood ratio test. The value of 1.0 used for $S_r - S_{\rm
min}$ corresponds to the 1-$\sigma$ confidence level for a single
interesting parameter ($r=1$).

\par

As clearly suggested by the imaging data, a considerably better fit is
provided by the elliptical isothermal-$\beta$ model.  The best-fit
elliptical model yields a value of $S_{\rm min}$ that is less than the
best-fit circular case by 35.5.  This is a highly significant
reduction in $S$ for the introduction of two additional parameters and
corresponds to a confidence level of greater than 5$\sigma$ for the
rejection of the circular model relative to the elliptical one.  Our
estimate of the absolute goodness-of-fit also indicates that this
model is a better fit than the previous one, but even this improved
$\chi^2$ of 117.5 can be formally rejected at about the 96.8\%
confidence level. Fitted parameters and errors for the elliptical
model are shown in Table 2. The center of the cluster is nearly
unchanged from the location fitted based on a spherical model.  The
cluster's ellipticity is quoted in terms of the ratio of major to
minor axis length, $e$, which appears in equations 6--9. The position
angle of the major axis is measured east of north. 
Figure 4a shows the radial profiles of the model and data azimuthally
averaged over quadrants and figure 4b shows the residuals. We note
that our numerical results are consistent with those of Neumann \&
B\"ohringer (1997).\footnote{$^2$}{We take this opportunity to correct
a statement by these authors identifying a disadvantage to our use of
maximum likelihood fits for X-ray images (near the end of \S 3.2 of
their paper).  In fact, our maximum likelihood fits have always
allowed for the determination of best-fit parameter values (and
confidence intervals) by locating the minimum of the figure-of-merit
function, $S$. Furthermore since our method is based on using the
appropriate Poisson distribution for the raw data, rather than
applying Gaussian filters to the data and assigning arbitrary error
values to blank pixels (as Neumann \& B\"ohringer do) it should result
in more robust parameter values and confidence intervals.}

\par\medskip
\centerline{3.3. {\it {Fits to the \SZ\ effect data}}}
The SZ effect data considerably undersample the structure of the
cluster (only a single north-south scan through \cl\ is available),
and thus cannot provide useful constraints on the model of the cluster
gas. For this reason we use the SZ effect data only to determine the
normalization of the SZ effect by fitting 
with structural models consistent with the X-ray imaging.
In the error analysis we were careful to include correlations among
parameters; in particular $\theta_C$, $\beta$, and the X-ray and
SZ-effect normalizations are highly correlated. 
\par
Using the parameters of the best-fit circular isothermal-$\beta$ model
(Table 1) we find a central SZ decrement of $\Delta T_{RJ}(0) = -1.20
\pm 0.19$~mK. This value includes the zero-level offset of $+70 \pm 43
\ \rm \mu K$ (as quoted in Sec.~2.2) and the error includes a
contribution from the uncertainty in the zero-level.  The $\chi^2$
associated with this fit is not particularly good, $\chi^2 = 10.5$ for
5 degrees of freedom, since the SZ effect data appear to be consistent
with a somewhat flatter central gas density distribution than the
X-ray data. The central SZ decrement in the absence of a
zero-level correction is smaller and corresponds to a central SZ
decrement of $\Delta T_{RJ}(0) \approx -1.0$ mK.  This fit is somewhat
worse: $\chi^2 = 13.0$ for 6 degrees of freedom.  The effect of the
zero-level offset is the largest systematic error in the \SZ\ effect
normalization and is particularly pernicious because it is a one-sided
error, and may therefore introduce a bias in the result for $H_0$.

\par

For the elliptical isothermal-$\beta$ model (Table 2) fits to the SZ
effect data result in nearly identical values for the central SZ
decrement $\Delta T_{RJ}(0) = -1.21 \pm 0.19$~mK, with a similar
goodness of fit to that for the circular model: $\chi^2 = 10.8$
for 5 degrees of freedom. 

\par\medskip
\centerline{3.4. {\it {Fits to the X-ray spectra}}}
\par

Our spectral fits are driven by two principal goals: (1) determination
of the mean temperature of the hot electrons, $kT_e$, in the plasma of
the intracluster medium, and (2) determination of the spectral
emissivity, $\Lambda_e$, of the gas as observed by the PSPC.  The
temperature is determined mainly by constraining the shape of the
X-ray continuum emission in the \asca\ data.  The spectral emissivity
is calculated from a standard thermal plasma code (we use the model of
Raymond \& Smith 1977; 1992 July 27 version) and depends on $kT_{e0}$, the
metallicity of the gas, and the column density of neutral hydrogen
($N_{\rm H}$) along the line of sight which causes absorption of soft
X-rays from the cluster. This latter quantity is most usefully
determined from the PSPC spectral data.  Clearly, therefore, a joint
spectral fit of the GIS and PSPC data will provide the most accurate
constraints on the relevant spectral parameters, leading in turn to
the most accurate values possible for the quantities of
interest.

The spectral fits were complicated in part by the X-ray emission from
QSO 0015+162 since this is not separated from the cluster emission by
the GIS.  The approach we took was to extract independent PSPC spectra
of QSO~0015+162 and \cl, and to fit appropriate spectral models to
each: a power-law for the AGN and a thermal plasma model for the
cluster. The sum of these models was required to fit the GIS data. In
total the spectral fits involved six free parameters: the temperature
and iron abundance of the cluster gas, the power-law index of the AGN,
the absorbing column density due to gas in our Galaxy (the same value
was used for both objects), and normalizations for both spectra. The
redshift of the cluster was fixed to the optically derived value,
$z=0.5455$.

The joint fit of the three datasets to the two spectral models is
formally acceptable with a minimum $\chi^2$ of 220.9 for 230 degrees
of freedom. Figure 2 shows the data, best fit models and residuals.
Table 3 presents numerical values for the fitted quantities.  The
cluster temperature corresponds to the value in the source-frame.  Our
values for $kT$ and iron abundance are consistent, within the errors,
with those quoted earlier by Yamashita (1994).  Note that the best-fit
column density $5.6\times 10^{20}$ atoms cm$^{-2}$ is only slightly
higher than the Galactic value of $4.1\times 10^{20}$ atoms cm$^{-2}$
(Stark \etal~1984). We define the emission measure of the cluster in
terms of the luminosity distance $D_L$ and the integral of the
electron and proton densities, $n_e$ and $n_p$, over the cluster
volume $V$ as $\int n_e n_p dV/(4\pi D_L^2)$.  The photon index of the
AGN power-law model, $dN/dE \sim E^{-\alpha_p}$ indicates a rather
steep spectrum that does not contribute greatly to the total X-ray
emission in the GIS 2--10 keV band.

We also investigated the sensitivity of the derived temperature to
uncertainty in background subtraction for the GIS data. Varying the
normalization of the background by $\pm6\%$ of the nominal value had
the effect of changing the best fit temperature by $^{-0.33}_{+0.35}$
keV.  These errors are included in the results of Table 3 (see
footnote), but are only about half the size of the statistical errors.

\par\medskip
\centerline{3.5. {\it {Relativistic Corrections to X-ray
Bremsstrahlung Emissivity}}}
\par

Use of the \SZ\ effect for distance determination to galaxy clusters
is based on our ability to calculate from fundamental physics accurate
expressions for the spectral intensity of X-ray bremsstrahlung
emission and the inverse Compton scattering distortions to the CMBR.
In most analyses to date nonrelativistic approximations have been
employed for these expressions.  For example the Raymond \&
Smith code mentioned above uses the Gaunt factor evaluated by Karzas
\& Latter (1961) in the nonrelativistic limit for the bremsstrahlung
component of the plasma emission.  Recently Rephaeli \& Yankovitch
(1997) have pointed out the importance of including relativistic
corrections (i.e., first order in $kT_e/mc^2$) for accurate work in
determining $H_0$ from rich galaxy clusters where $kT_e$ can be as
large as 15 keV. In the following we employ their expressions for the
relativistic corrections to the intensity change of the CMBR (which
agree with other calculations, e.g., Birkinshaw 1998), but have found
an error in their relativistic X-ray bremsstrahlung emissivity which
we detail below.

Although we refer to these as ``relativistic'' corrections, in fact,
the results that we and Rephaeli \& Yankovitch both use for X-ray
bremsstrahlung come from Gould (1980). These calculations use the
nonrelativistic Born approximation as the principle term for the
electron-ion bremsstrahlung cross section and then include first-order
corrections from relativistic effects (both to the thermal electron
velocity distribution function and the electron-ion bremsstrahlung
cross section), emission from electron-electron bremsstrahlung, and a
first order correction to the Born approximation, which is most
important at lower temperatures.  These modifications are each of
order 10\% and, when applied, are expected to result in a formula for
thermal bremsstrahlung emission that is accurate to 1\% for a $\sim$10
keV plasma.  Gould (1980) presents results for both the total
energy-loss rate (integrated over frequency) and the spectral
emissivity of thermal bremsstrahlung.

Rephaeli \& Yankovitch start from Gould's formula for the total
energy-loss rate which they note is significantly larger than
expressions for relativistic corrections to bremsstrahlung published
in standard texts (e.g., Rybicki \& Lightman 1979).  We have
identified a 
misprint\footnote{$^3$}{In the brackets in equation 41 of Gould (1980),
the value 8 in the denominator of the second term should be replaced
with the value 24. This follows directly from equation 22.}
in Gould (1980) that accounts for most of this difference.  However,
considering this issue further we find that we disagree with the use
of the integrated bremsstrahlung energy-loss rate since what matters
is the effects of the relativistic corrections on (1) the calculated
value of $\Lambda_e$, which, as discussed above, comes from
integrating the thermal bremsstrahlung emissivity function over the,
relatively narrow, redshifted \rosat\ energy band, and (2) the fitted
value of $kT_e$ from X-ray spectra.  Addressing both of these issues
requires use of the spectral emissivity formula and so this is what we
have chosen to implement.  As a check we have verified that the
Gould's spectral emissivity formula (eqn.~43), when integrated over
frequency, agrees with the (corrected) total energy-loss rate formula.
 
Rephaeli \& Yankovitch argue that the effect of relativistic
corrections on the fitted value of $kT_e$ should be small since it is
the exponential factor that dominates the shape of a bremsstrahlung
spectrum and hence drives the spectral fits.  Work we have done confirms
that suspicion.  In the course of analyzing the \ginga\ spectrum of
the Coma cluster (Hughes \etal\ 1993), which covered the 2--20 keV band,
we used both the Karzas \& Latter (1961) and Gould (1980) calculations
to derive the cluster temperature.  The $kT_e$ values obtained: $8.07
\pm 0.09$ and  $7.96 \pm 0.09$, respectively, differ by only
$\sim$1\%, and clearly establish that the inclusion of relativistic
corrections to X-ray bremsstrahlung emissivity do not significantly change
fitted temperature values.

As for the numerical value of $\Lambda_e$, we used the results in
Table 3 to calculate the spectral emissivity of the hot plasma in \cl\
over the identical PSPC X-ray band used for the imaging analysis.  For
the best-fit spectral values and including relativistic corrections we
find $\Lambda_e = 2.88\times 10^{-13}$ PSPC ct s$^{-1}$ cm$^{-5}$.
(Note that this relativistically-corrected value for $\Lambda_e$ is
only 1.048 times larger than the nonrelativistic value.)  The
variation of $\Lambda_e$ is less than 2.5\% over the entire range of
parameter values bounded by the errors quoted in Table 3.

\bigskip
\centerline {{\bf 4. THE VALUE OF THE HUBBLE CONSTANT}}
\par\medskip
\centerline {4.1. {\it{Effect of Different Cosmologies}}}
\par

In our previous work on the relatively nearby clusters Abell 665 (BHA)
and Abell 2218 (Birkinshaw \& Hughes 1994) we presented values of
$H_0$ assuming a Friedmann cosmology and a value of 0 for the
deceleration parameter, $q_0$.  At the redshifts of these clusters ($z
\sim$ 0.17 -- 0.18), the dependence of $H_0$ on the values of the
other cosmological parameters is not strong.  For example, the
assumption of an Einstein-de Sitter universe ($q_0 = 0.5$) would have
changed our derived values of $H_0$ by merely 3\%.  The situation is
different for the more distant cluster \cl, where distance estimates
depend significantly not only on the mean density of the universe, but
on the value of the cosmological constant (Kobayashi, Sasaki, \& Suto
1996).

\par

Although our results for \cl\ are not accurate enough to place
significant observational constraints on the density of the universe
or the cosmological constant, we nonetheless wish to present our
results in the context of the currently reasonable range of acceptable
cosmologies. To this end we write our equation for $H_0$ as
$$H_0 = {cz \over D_A} {1+z/2 \over (1+z)^2}
f(z,\Omega_0,\lambda_0), \eqno(11)$$ 
where $f(z,\Omega_0,\lambda_0)$ contains
all the functional dependence on the density parameter $\Omega_0$ and
the dimensionless cosmological constant $\lambda_0$.  Our definition
of $f$ is such that for $\Omega_0 = 0$ and $\lambda_0=0$, $f=1$. In the
case of Friedmann universes (\ie, $\lambda_0=0$), $f$ has an
analytical closed form expression (see equation 15.3.24 in Weinberg
1972), although in the more general case with a non-vanishing
cosmological constant an integral over redshift must be carried out
(see Carroll, Press, \& Turner 1992).  In Figure 5 we present
curves of $f$ versus $z$ for several representative values of
$\Omega_0$ and $\lambda_0$.  The two extreme curves correspond to flat
universes where $\Omega_0 + \lambda_0 = 1$, as favored by inflationary
models of the early Universe.

\par

In the following we quote $H_0$ values for \cl\ assuming $\Omega_0=0.2,
\lambda_0=0$, for which $f = 0.97$ at the cluster's redshift 
($z=0.5455$). For the curves shown in Fig.~5 $f$ varies from 0.87 to
1.10 at the same redshift. 

\par\medskip
\centerline {4.2. {\it{Basic Error Analysis}}}
\par

The value of the Hubble constant derived for the best-fit circular
isothermal-$\beta$ model (Table 1) using the best-fit spectral
parameters (Table 3) is 47.2 \kmsMpc.  The error budget is presented
in Table 4 (values are at the 68.3\%, or $1\sigma$, confidence
level).  The errors are dominated by the measurements of temperature
and the SZ-effect normalization.  Uncertainty in metallicity and
column density enter into the error budget only through their effect
on $\Lambda_e$.  Although the uncertainty in the metallicity of the
gas is large, at the relatively high temperature of the cluster there
is little emission from metals in the \rosat\ band and so the
variation in $\Lambda_e$ is small. The variation in $\Lambda_e$ due to
the error range in column density is also small: the calculated X-ray
absorption at an energy of 1 keV varies by only about $\pm1$\% over
the allowed range of $N_{\rm H}$.  Note that only a single value is
presented for the uncertainty in $H_0$ due to the structural
parameters of the circular model ($b_{X}(0)$, $\beta$, and $\theta_C$).
Since these parameter values are correlated, as mentioned above, we
determined their effect by calculating trial values of $H_0$ over the
three-dimensional 68.3\% confidence level error surface (specifically
for $S-S_{\rm min} = 3.53$).  Clearly the error due to statistical
uncertainty in the parameter values associated with fits to this
particular model are quite small.  However, as we see below, these
errors are entirely dominated by the uncertainties in deprojection of
the elliptical model fits.

\par

In addition to the quantities shown in Table 4 we also have
investigated systematic errors due to overall flux or brightness
temperature calibration uncertainties.  The absolute flux of the
cluster is probably uncertain by $\pm 10$\% due to residual
uncertainties in the overall effective area calibration of
\rosat, which results in an uncertainty in $H_0$ of $\pm 4.7$ \kmsMpc.
Likewise, the overall SZ-effect normalization, $\Delta T_{RJ}(0)$ is
uncertain by $\pm 8$\% due to uncertainty in the efficiency of the 40-m
telescope for an error in $H_0$ of $^{-6.8}_{+8.5}$ \kmsMpc.
Combining all the errors in quadrature we obtain
$$H_0 = 47^{+23}_{-16}\, \rm km\, s^{-1}\, Mpc^{-1}. \eqno(12)$$
\par\noindent
If the effect of a zero-level offset to the SZ-effect data ($+70\, \rm
\mu K$) were not included in result (12), the derived value would be 
increased by $21$ \kmsMpc.

The breakdown of errors in Table 4 indicate where improvement of the
observations should be focussed.  Clearly a longer observation with
\asca\ to reduce the statistical error in temperature would be
extremely valuable. Indeed a long follow-up observation of \cl\ by \asca\
was carried out last year but the data are not generally available
yet.  The other major, and in fact dominant, source of error is in
the SZ-effect normalization which also enters quadratically in eqn.~5.
Interferometric observations of the SZ-effect from \cl\ and other
clusters have been reported recently (Carlstrom, Joy, \& Grego 1996).
These data are significantly less susceptible to errors from
zero-level offsets and are of considerably higher sensitivity as
compared to the data from the 40-m telescope, which in combination
should greatly reduce the overall errors on $H_0$. In a future article
we will be comparing our models for the cluster atmosphere to these
interferometric SZ-effect data to derive a more precise value of the
Hubble constant.

\par\medskip
\centerline {4.3. {\it{Effects of Unknown Geometry}}}
\par

The accuracy in any measurement of $H_0$ from a single cluster,
however, is ultimately limited by uncertainties in the unknown
three-dimensional geometry of the target cluster.  For example, the
isothermal-$\beta$ model we use in our analysis is unbounded, i.e.,
the gas distribution extends to infinite radial extent.  Although this
is clearly an approximation to the situation of a real cluster, it
turns out to be a rather good one. In cases that have been
well-observed, like the Coma Cluster (Briel, Henry, \& B\"ohringer
1992), the isothermal-$\beta$ model provides an excellent description
of the X-ray surface brightness profile out to roughly 10 core
radii. Nevertheless it is important to consider the effect that a
radial truncation in the cluster gas distribution might introduce in
the determination of $H_0$ for any particular cluster.  

The radially averaged X-ray surface brightness of \cl\ can be
confidently traced to approximately 5$^\prime$, or roughly
7$\theta_C$, where the cluster emission is $\sim$25\% of the local
background rate. We implement a truncated model by modifying the
isothermal-$\beta$ model form factor $f_n$ used to derive equations
(3) and (4) so that $f_n = 0$ for $r > r_{\rm max}$.  Using the
truncation radius determined from the cluster's maximum angular
extent, we fit for new values of $N_X$, $\beta$, and $\theta_C$ from
the X-ray image and for a new value of $N_{RJ}$ from the SZ effect
scan data (using the new best-fit $\beta$ and $\theta_C$ values),
assuming a circular isothermal-$\beta$ model.  The best fit values of
$\beta = 0.69$ and $\theta_C = 0\arcmin.64$ are somewhat smaller (and
the quality of the fit is worse) than the untruncated model.  Carrying
the new normalization factors through to a value for the Hubble
constant we find $H_0 = 50.2$ \kmsMpc, some 6\% higher than our best
estimate based on the untruncated  model.
This difference is given by nearly equal contributions from a 3\%
increase in $N_X$ and a 1.5\% decrease in $N_{RJ}$.  This truncation
uncertainty in $H_0$ is somewhat larger than Holzapfel et al.~(1997)
found in their work on $H_0$ using cluster Abell 2163 --- an 
increase in the derived value of $H_0$ of only 1\% due to truncation.
This is at least partially due to the very large radius
($\sim$18$^\prime$) to which the X-ray emission of Abell 2163 
extends and over which the isothermal-$\beta$ model is an excellent
fit.

\par

An even larger source of uncertainty in determining $H_0$ comes from
the elliptical morphology of galaxy clusters.  In the following we
quantify this uncertainty numerically using our axisymmetric
ellipsoidal models for the gas distribution of \cl. Employing equations
(6)--(9) and the best-fit values in Tables 2 and 3, we arrive at the
values in Table 5 for $H_0$ as a function of line-of-sight inclination
angle.  Even if the symmetry axis of the cluster is assumed to lie in
the plane of the sky (inclination angle of $i=90^\circ$), then whether
the cluster is oblate or prolate introduces an uncertainty of
approximately $\pm$8\% in $H_0$ compared to the circular model.  In
fact it is easy to see that for $i=90^\circ$ the ratio of Hubble
constants is $H_0({\rm oblate})/H_0({\rm prolate}) = e$, which comes
directly from the different line-of-sight depths through the cluster
under the two assumptions.

\par

Results for different assumed values of the inclination angle of the
symmetry axis of the cluster to the line-of-sight are shown
graphically in Figure 6 and some numerical values are given in Table
5.  We have chosen to plot $H_0$ versus the intrinsic major/minor axis
ratio of the cluster, rather than versus inclination angle, although
the results are equivalent.  Clearly the minimum intrinsic axis ratio
is obtained when the cluster's symmetry axis lies in the plane of the
sky, $i=90^\circ$, which corresponds to the left side of Figure 6.  As
the symmetry axis of the ellipsoid is allowed to vary toward the
line-of-sight ($i=0^\circ$), then the intrinsic major/minor axis ratio
grows increasingly larger, as does the uncertainty on $H_0$.  This
appears in Figure~6 as the increasing difference between the prolate
and oblate curves as the abscissa increases.

\par

In the absence of other information about the structure of galaxy
clusters, the error in $H_0$ from these geometric effects can be quite
large.  However, a reasonable method for bounding the $H_0$ error is
to bound the range of observed, projected ellipticities of other
clusters. Mohr \etal\ (1995) analyzed a sample of galaxy clusters,
fitting elliptical isothermal-$\beta$ models to their X-ray images,
and found a mean value for the observed major/minor axis ratio of
$1.25\pm0.19$ from the sample.  Of course this is not a measurement of
the intrinsic axial ratio of clusters, but it does indicate that
highly elliptical clusters tend to be rare.  Guided by these results
we make the assumption that the intrinsic value of the major/minor
axis ratio of an individual cluster is unlikely to be greater than 1.5
(this value is shown as the vertical dashed line in Figure 6) and find
that the range of $H_0$ is then bounded to 40 -- 51 \kmsMpc\ if the
cluster is oblate and 43 -- 55 \kmsMpc\ if the cluster is prolate.
This total range is $\pm$17\% when expressed as a fractional error.

\par

How might these results generalize to clusters with other values of
observed ellipticity?  Let us consider two extreme cases.  The first
case assumes that the cluster appears circularly symmetric. This might
indicate that the cluster is indeed spherically symmetric, but it
could also mean that the cluster is oblate or prolate with a symmetry
axis lying directly along the line-of-sight.  One needs to consider
the latter situation in order to assess honestly the systematic error
due to unknown geometry. Doing that we find the fractional error in
$H_0$ to be $\pm$38\%.  In the second case we assume that the cluster
has an observed major/minor axis ratio of 1.5, i.e., the cluster has
the maximum intrinsic ellipticity and the symmetry axis lies in the
plane of the sky.  Now we find that the conservative error estimate,
assuming oblate and prolate geometries, yields a fractional error in
$H_0$ of $\pm$20\%.

More definitive information about the intrinsic structures of clusters
could significantly reduce these systematic errors.  For example, if
we could be certain that clusters were nearly always prolate, then the
fractional errors associated with the two extreme cases presented
above would be reduced to $\pm$20\% and 0\%, respectively.  Until more
theoretical and observational studies give us such insights into the
nature of galaxy clusters, the simplified arguments presented in the
preceding paragraph should be kept in mind when considering the total
error budget associated with X-ray/SZ-effect distance measurements.

\par\medskip
\centerline {4.4. {\it{Comparison with Other SZ-derived Values for the
Hubble Constant}}} 
\par

There are seven other galaxy clusters in addition to \cl\ with
published measurements of $H_0$ from the SZ effect based on various
observational techniques. In the rest of this section we compare these
results. When necessary and as indicated below, we have applied the
appropriate relativistic corrections for the intensity of X-ray
bremsstrahlung emissivity and the inverse Compton scattering
distortions to the CMBR using the techniques described above.  We have
made sure that the various $H_0$ values were determined in a
consistent fashion with the same assumptions about the cluster
structure and temperature distribution.  With the exception of the
Coma Cluster for which a nonisothermal temperature distribution was
used, the values we present in Table 6 were determined assuming
spherical symmetry; an isothermal-$\beta$ model for the gas density
distribution; isothermal, unclumped gas; and $\Omega_0=0.2,
\lambda_0=0$.   Uncertainties are nearly entirely observational and
are given at the 1 $\sigma$ confidence level.

Myers \etal~(1997) present results of SZ effect observations in four
nearby clusters of galaxies from single dish measurements at 32 GHz
with the OVRO 5.5-m telescope. In their analysis these authors include
relativistic corrections for the SZ effect, but not for the X-ray
bremsstrahlung emission, since their X-ray results come largely from
the literature. For the mean cluster temperatures determined by
\ginga, we find that the relativistically-corrected $H_0$ values are
smaller than the Myers \etal\ values by 5.2\% (A2256,
A478), 5.4\% (A2142), and 5.5\% (Coma).  The Myers \etal\ values also
need to be increased slightly (adjustments range from 0.5\% for
Coma to 1.8\% for A2142) for consistency with our different assumed
cosmology.

The SZ detection of the Coma cluster was first reported by Herbig
\etal\ (1995), whose analysis, as mentioned above, incorporated a
nonisothermal temperature distribution as derived from various
X-ray data sets by Hughes, Gorenstein, \& Fabricant (1988).  The
general shape of the radial temperature variation of Coma is both
strongly motivated and observationally secure (see Hughes 1997 for a
summary of recent results and models), so Herbig \etal 's use of it is
appropriate. Unfortunately, these authors do not quote an $H_0$ value
for an isothermal temperature distribution, and it is not possible to
estimate one from their article, so the sensitivity of the derived
value of $H_0$ to the assumption of nonisothermality for the Coma
Cluster cannot be assessed.  We do not use Coma's $H_0$ value for the
ensemble average in Table 6, but consider it later as discussed below. 

Our published value of $H_0$ from the SZ effect of A2218 ($65\pm25$
\kmsMpc, Birkinshaw \& Hughes 1994) has been adjusted for
relativistic corrections to the SZ effect (reduced by 4\%) and X-ray
bremsstrahlung emissivity (reduced by 5.2\%) and for consistency with
our different assumed cosmology (reduced by 1\%).  These same three
corrections were applied to the result from Jones (1995) of $H_0 =
38^{+18}_{-16}$ \kmsMpc, which were based on interferometric SZ maps
from the Ryle Telescope observing at 15 GHz and \rosat\ PSPC X-ray
images.

The result on A665 quoted in Table 6 comes from a detailed analysis of
a deep \rosat\ PSPC image (Hughes \& Birkinshaw 1998) and uses the
OVRO 40-m telescope SZ effect data from Birkinshaw \etal\ (1991). It
has been corrected for relativistic effects and has been computed for
the assumed cosmological parameters.

The SZ effect measurements of A2163 from an infrared bolometer array
observing at 2.1 mm (SuZIE, the \SZ\ Infrared Experiment) and the
\rosat\ PSPC X-ray data were analyzed in a comprehensive article by
Holzapfel \etal~(1997).  We have selected their value of $H_0$ derived
for an isothermal temperature distribution ($59.6^{+40.7}_{-22.6}$
\kmsMpc) and applied relativistic corrections for the X-ray
bremsstrahlung emissivity (which reduced their value by 7.7\%) and for
our different assumed cosmology (which increased their value by 4\%).

The ensemble average for $H_0 = 47.1 \pm 6.8 \ \rm km \, s^{-1} \,
Mpc^{-1}$ was computed by minimizing $\chi^2$ assuming the eight
measurements (excluding Coma) are independent data points, and
including the asymmetry of the error bars on the individual
measurements. The assumption of independence is not strictly correct
since the several observations share systematic errors and the two
values from A2218 share the same measurement of the cluster
temperature from \ginga. Nevertheless this procedure gives a mean
value that is comparable to the weighted (45 \kmsMpc) or unweighted
(49 \kmsMpc) average of the eight measurements.  None of the
individual results differs by more than 1.5 $\sigma$ from the ensemble
average.

\par\medskip
\centerline {4.5. {\it{Systematic Errors and Biases}}}
\par

The excellent agreement among the various $H_0$ measurements quoted
above notwithstanding, any serious attempt at determining cosmological
parameters from the SZ effect demands a careful and intensive analysis
of systematic effects that might be sources of error. We make an
important distinction between systematic errors that may introduce a
bias in an ensemble average value versus errors that may introduce a
random uncertainty.  The dominant sources of biases are calibration
errors in the flux scale of X-ray measurements or the brightness
temperature of the SZ effect measurements, large scale temperature
gradients in cluster atmospheres, clumping of cluster gas, and a
poorly selected sample of clusters.  Random errors include primarily
any motion of a cluster with respect to the Hubble flow and cluster
morphology, such as the effects of ellipticity as studied in detail
above. Clearly the effects of biases are particularly pernicious since
they cannot be eliminated by increasing the size of the cluster
sample.  Rather, their effect on the derived value of $H_0$ must be
carried along as an additional error term.  In the following we draw
upon work of others as well as our own to quantify these various
systematic errors.

Birkinshaw \& Hughes (1994) and Holzapfel \etal\ (1997) allowed for
large scale radial temperature gradients when analyzing the SZ effect
and X-ray data for A2218 and A2163, respectively. Both groups found
that, for temperature profiles that fell with radius (as is the case
for the temperature profile of the Coma cluster), the value of $H_0$
derived under an isothermal assumption would underestimate the true
$H_0$ value by 20\%--30\%. Inagaki, Suginohara, \& Sato (1995) came to
a similar conclusion based on studies of simulated clusters with
temperature distributions that fell with radius.  Our own comparison
of $H_0$ measurements in Table 6 also bears this out to some extent.
Specifically the best-fit value of $H_0$ from the Coma Cluster, for
which a nonisothermal temperature distribution was used, is about 36\%
greater than the ensemble average of the other clusters, for which the
gas was assumed to be isothermal.

If cluster gas is clumped, then X-ray emissivity will be increased
relative to SZ by a factor greater than unity (BHA).  In this case the
value of $H_0$ derived assuming an unclumped gas distribution will be an
upper limit to the true $H_0$ value. Holzapfel \etal\ (1997) used
X-ray spectral fits to constrain the amount of isobaric clumping in
A2163 and found that a reduction in $H_0$ of only $\sim$10\% from the
unclumped case was allowed.

The peculiar motion of clusters relative to the Hubble flow introduces
an additional distortion to the CMBR spectrum usually referred to as
the ``kinematic'' SZ effect.  For a cluster with a peculiar velocity
of $1000\,\rm km\,s^{-1}$ and temperature of $10\,\rm keV$ the
strength of the kinematic SZ effect would be $\sim$9\% of the thermal
effect in the Rayleigh-Jeans portion of the CMBR spectrum
(BHA). Recently Watkins (1997) presented observational evidence that
argues for a low 1-D RMS peculiar velocity of clusters, $\sigma_v \sim
300$ km s$^{-1}$.  Assuming this value and recalling that the SZ
effect intensity enters as a square in the equation determining $H_0$,
we determine that the kinematic SZ effect could introduce up to a
$\sim$$\pm$8\% correction in the measurement of $H_0$ from any
individual cluster (for $kT_{e0} = 7$ keV).  It is important to note that
this effect produces a spectral signature that is different from the
thermal effect and so, in principle, is amenable to observation. In
practice, current results are not sensitive enough to set significant
constraints; the SuZIE observations of A2163 constrain its 1-D
peculiar velocity to be less than 1500 km s$^{-1}$.  Peculiar
velocities are unlikely to be correlated for clusters that are widely
distributed in redshift and position, so this effect would result in
an additional random uncertainty in $H_0$ for any single cluster.

The effect of cluster ellipticity that we have studied in some detail
in this article tends to introduce a random uncertainty of about $\pm
20\%$ in $H_0$ for an average cluster. However, to ensure that these
effects of unknown geometry and arbitrary inclination are uncorrelated
from cluster to cluster, it is essential that the cluster sample for
determining $H_0$ be selected properly.  For example, as pointed out
by BHA, it is important that clusters {\it not} be selected based on
the strength of their SZ effect signal or central X-ray surface
brightness, since this would result naturally in a bias toward prolate
clusters with their long axes aligned to the line-of-sight.  As figure
6 clearly shows, this bias would cause the derived value of $H_0$ to
be an underestimate of the true value.  The four low redshift objects
in Table 6 are part of an X-ray flux-limited sample of nearby clusters
that are being studied in the SZ effect (Myers \etal\ 1997).  The four
other clusters are part of a moderate redshift sample selected on the
strength of their integrated X-ray flux from surveys by the {\it
Einstein Observatory} or \rosat\ that we and our collaborators have
been observing in the X-rays and SZ effect.  Although the entire
sample in each case should be relatively unbiased, it is not clear
that the same may be said about the subsamples presented here.  A
definitive answer awaits the final analysis of the entire sample.

We take account of the systematic errors mentioned above by quoting a
range in $H_0$ that includes a bias of +30\% from large scale
temperature gradients and one of $-$10\% from clumping in the cluster
gas.  Additional random systematic errors on the ensemble average are
$\pm3$\% from peculiar motion and $\pm$7\% from unknown
geometry/inclination, which the reader will note have been reduced
from the values given above by $1/\sqrt{N}$ where $N=8$ is the number
of clusters in the sample. These random errors are then
root-sum-squared with the 14\% error on the ensemble average quoted in
Table 6.  This yields a best estimate of the Hubble constant from the
\SZ\ effect of

$$ H_0 = 42 - 61 \,\,\rm km\, s^{-1}\, Mpc^{-1} \,\, \pm 16\%. \eqno(13)$$

\bigskip
\centerline {{\bf 5. CONCLUSIONS}}

In this article we have arrived at the following conclusions.

\item{(1)} The structure of the intracluster medium of \cl\ is well
described by a gas distribution that produces in projection an
elliptical isothermal-$\beta$ model with $\beta\approx 0.74$,
$\theta_C\approx 0.75$ arcmin (along the major axis), and
observed major/minor axis ratio $\approx 1.18$ with the major axis
lying at position angle $\approx 51^\circ$ measured east of north.

\item{(2)} If the cluster is assumed to be spherically-symmetric and
isothermal, then we derive a value for the Hubble constant of $H_0 =
47 ^{+23}_{-15}\, \rm km\, s^{-1}\, Mpc^{-1}$.  The error includes
random uncertainties from fits to the model parameters and systematic
uncertainties in background subtraction (for \asca\ spectroscopy), the
overall \rosat\ X-ray flux calibration, and the brightness temperature
scale of the OVRO 40-m telescope. An additional, one-sided error of
$+21\, \rm km\, s^{-1}\, Mpc^{-1}$ comes from the zero level
uncertainty.

\item{(3)} We have quantified the error in $H_0$ from unknown
geometry and line-of-sight inclination angle effects, under the
assumption that \cl\ is an axisymmetric ellipsoidal system.  The $H_0$
error can be bounded if we assume that the intrinsic major/minor axis ratio of
the cluster is less than 1.5. In this case we find (not including
observational errors)
$$ 
  \eqalign{
     \rm Oblate:  &\quad H_0 = 40 - 51 \quad \rm km \, s^{-1} \, Mpc^{-1} \cr
     \rm Prolate: &\quad H_0 = 43 - 55 \quad \rm km \, s^{-1} \, Mpc^{-1}. \cr
}
$$
These ranges represent an irreducible uncertainty in the value of
$H_0$ as determined from this single cluster, and indicate the sizes
of the errors that would be caused by reasonable triaxial models for
the cluster. In the more general case of an arbitrary observed
ellipticity we find that the fractional error associated with $H_0$ is
probably not much smaller than $\pm15$\% and in the worst case may be
as large as $\pm38$\%.

\item{(4)} At the redshift of \cl, $z=0.5455$, the effects of the
cosmological parameters on the derived value of $H_0$ can be of order
10\%--20\%. The results given above are for a Friedmann Universe with
$q_0 = 0.1$. Turning the problem around and assuming that the results
summarized in item (3) above can be applied to other clusters, in
order to place significant constraints on the cosmological parameters
using the SZ effect it will be necessary to utilize a carefully
selected sample of $\sim$20 or more clusters at redshifts beyond
$\sim$0.2.

\item{(5)} When the result for \cl\ given here is compared with other
determinations of $H_0$ from the SZ effect (see Table 6), a consistent
ensemble value of $47 \pm 7 \, \rm km\, s^{-1}\, Mpc^{-1}$ is obtained
under the assumption of a spherical, unclumped, isothermal cluster
atmosphere. This mean value depends only slightly on the assumed
cosmology varying by $^{+2.6}_{-1.4}$ \kmsMpc\ for the extreme curves
shown in Figure 5.  Inclusion of likely systematic errors increases
the allowed range of $H_0$ to 42 -- 61 \kmsMpc\ with a random error of
$\pm 16$\%. It is reassuring that this result is consistent with other
independent measures of $H_0$ such as those from using Type Ia
supernovae (SNe) as standard candles: 62--67 \kmsMpc\ (Hamuy \etal\
1995) and $67 \pm 7$ \kmsMpc\ (Riess, Press, \& Kirshner 1996); the
expanding photosphere method on Type II SNe $73\pm 7$ \kmsMpc\
(Schmidt \etal\ 1994); or from measurements of time delays in
gravitational lenses: $64 \pm 13$ \kmsMpc\ (Kundi\'c \etal\ 1997) and
$41 - 84$ \kmsMpc\ (Schechter \etal\ 1997).

\item{(6)} The dominant source of error in the determination of $H_0$
from the SZ effect is the possible existence of large scale radial
temperature gradients in cluster atmospheres.  It is interesting to
note that the presence of negative radial gradients, similar to the
kind observed in the Coma cluster, is roughly what is needed to bring
the low values of $H_0$ determined assuming isothermal atmospheres
into agreement with the results from supernovae mentioned above.  The
temperature distribution of clusters can be studied observationally by
comparing interferometric maps of the SZ effect (which sample
$n_eT_e$) with X-ray images (which effectively sample $n_e^2$) or
through direct spatially resolved X-ray spectral measurements with the
upcoming satellites \axaf\ and \xmm. Over the next few years we expect
to make substantial progress in resolving this key issue.

\par\medskip
\centerline{Acknowledgements}

\par

This research has made use of data obtained through the High Energy
Astrophysics Science Archive Research Center Online Service, provided
by the NASA/Goddard Space Flight Center. JPH was partially supported
by NASA \rosat\ Grant NAG5-2156 and NASA Long Term Space Astrophysics
Program Grant NAG5-3432. MB was partially supported by NASA grants
NAGW-3825 and NAG5-2415, NASA contract NAS8-39073, and a PPARC grant.

\vfill\eject

\centerline{\bf{REFERENCES}}
\smallline
\hangindent=\parindent
Birkinshaw, M.~1979, \mnras, 187, 847
\smallline
\hangindent=\parindent
Birkinshaw, M.~1998, Physics Reports, in press.
\smallline
\hangindent=\parindent
Birkinshaw, M., Gull, S.F., Hardebeck, H.E. \& Moffet, A.T., 1997,
 \apj, submitted.
\smallline
\hangindent=\parindent
Birkinshaw, M., Hughes, J.~P., \& Arnaud, K.~A.~1991, \apj, 379, 466 (BHA)
\smallline
\hangindent=\parindent
Birkinshaw, M., \& Hughes, J.~P.~1994, \apj, 420, 33
\smallline
\hangindent=\parindent
Briel, U.~G., Henry, J.~P., and B\"ohringer, H.~1992, \aa, 259, 131
\smallline
\hangindent=\parindent
Carlberg, R.~G., Yee, H.~K.~C., Ellingson, E., Abraham, R., Gravel,
P., Morris, S., \& Pritchet, C.~J.~1996, \apj, 462, 32
\smallline
\hangindent=\parindent
Carlstrom, J.~E., Joy, M., \& Grego, L.~1996, \apj, 456, L75
\smallline
\hangindent=\parindent
Carroll, S.~M., Press, W.~H., \& Turner, E.~L.~1992, \araa, 30, 499
\smallline
\hangindent=\parindent
Cavaliere, A., Danese, L., \& DeZotti, G.~1979, \aa, 75, 322
\smallline
\hangindent=\parindent
Dressler, A., \& Gunn, J.~E.~1992, \apjs, 78, 1
\smallline
\hangindent=\parindent
Fabricant, D., Rybicki, G., \& Gorenstein, P.~1984, \apj, 286, 186
\smallline
Fixsen, D.~J., Cheng, E.~S., Gales, J.~M., Mather, J.~C., Shafer, R.~A. \&
Wright, E.~L.~1996, \apj, 473, 576
\hangindent=\parindent
\smallline
\hangindent=\parindent
Gould, R.~J.~1980, \apj, 238, 1026; erratum 243, 677 (1981)
\smallline
\hangindent=\parindent
Gunn, J.~E.~1978, in Observational Cosmology, eds.~A.~Maeder,
L.~Martinet, \& G.~Tammann (Geneva: Geneva Observatory)
\smallline
\hangindent=\parindent
Hamuy, M., Phillips, M.~M., Maza, J., Suntzeff, N.~B., Schommer,
R., \& Aviles, A.\ 1995, \aj, 109, 1
\smallline
\hangindent=\parindent
Hasinger, G., Turner, T.~J., George, I.~M., \&  Boese, G.~1992,
Legacy, 2, 77 
%
\smallline
\hangindent=\parindent
Herbig, T., Lawrence, C.~R., Readhead, A.~C.~S., \& Gulkis,
S.~1995, \apj, 449, L5
\smallline
\hangindent=\parindent
Holzapfel, W.~L., Arnaud, M., Ade, P.~A.~R., Church, S.~E., 
Fischer, M.~L., Mauskopf, P.~D., Rephaeli, Y., Wilbanks, T.~M., \&
Lange, A.~E.~1997, \apj, 480, 449.
\smallline
\hangindent=\parindent
Hughes, J.~P.~1997, in the Proceedings of the conference {\it A New
Vision of an Old Cluster: Untangling Coma Berenices} (held June 17-20,
Marseille, France), in press (astro-ph/9709272).
\smallline
\hangindent=\parindent
Hughes, J.~P., \& Birkinshaw, M.~1998, \apj, in preparation
\smallline
\hangindent=\parindent
Hughes, J.~P., Birkinshaw, M., \& Huchra, J.~P.~1995, \apj, 448, L93
\smallline
\hangindent=\parindent
Hughes, J.~P., Butcher, J.~A., Stewart, G.~C., \& Tanaka, Y.~1993,
\apj, 404, 611
\smallline
\hangindent=\parindent
Hughes, J.~P., Gorenstein, P., and Fabricant, D.~1988, \apj,  329, 82
\smallline
\hangindent=\parindent
Inagaki, Y., Suginohara, T., \& Suto, Y.~1995, PASJ, 47, 411
\smallline
\hangindent=\parindent
Jones, M.~1995, Astrophys.~Lett.~\& Comm., 32, 1-6, 347
\smallline
\hangindent=\parindent
Karzas, W., and Latter, R.~1961, \apjs,  6, 167
\smallline
\hangindent=\parindent
Kendall, M., \& Stuart, A.~1979, The Advanced Theory of Statistics,
Volume 2 (New York: Macmillan), 246ff
\smallline
\hangindent=\parindent
Kobayashi, S., Sasaki, S., \& Suto, Y.~1996, \pasj, 48, 107
\smallline
\hangindent=\parindent
Kundi\'c, T., \etal\ 1997, \apj, 482, 75
\smallline
\hangindent=\parindent
Madore, B.~F., \etal\ 1996, BAAS, 28, 1420
\smallline
\hangindent=\parindent
Margon, B., Downes, R.~A., \& Spinrad, H.~1983, \nature, 301, 221
\smallline
\hangindent=\parindent
Moffet, A.~T., \& Birkinshaw, M.~1989, \aj, 98, 1148
\smallline
\hangindent=\parindent
Mohr, J.~J., Evrard, A.~E., Fabricant, D.~G., \& Geller, M.~J.~1995,
\apj, 447, 8
\smallline
\hangindent=\parindent
Myers, S.\ T., Baker, J.~E., Readhead, A.~C.~S., \& Herbig, T.~1997,
\apj, 485, 1
\smallline
\hangindent=\parindent
Ohashi, T., \etal\ 1996, \pasj, 48, 157
\smallline
\hangindent=\parindent
Perlmutter, S., \etal~1997, \apj, 483, 565
\smallline
\hangindent=\parindent
Pfeffermann, E., \etal\ 1986, Proc.~SPIE Int.~Opt.~Eng., 733, 519
\smallline
\hangindent=\parindent
Press, W.~H., Flannery, B.~P., Teukolsky, S.~A., \& Vetterling,
W.~T.~1985, Numerical Recipes, Ed.~1 (Cambridge: Cambridge University
Press), 289
\smallline
\hangindent=\parindent
Rephaeli, Y., \& Yankovitch, D.~1997, \apj, 481, L55
\smallline
\hangindent=\parindent
Riess, A.~G., Press, W.~H., \& Kirshner, R.\ P.\ 1996, \apj, 473, 88.
\smallline
\hangindent=\parindent
Rybicki, G.~B., \& Lightman, A.~P.~1979, Radiative Processes in
Astrophysics (New York: Wiley), p.~165
\smallline
\hangindent=\parindent
Schechter, P.~L., \etal\ 1997, \apj, 475, L85
\smallline
\hangindent=\parindent
Schmidt, B.~P., \etal~1994, \apj, 432, 42
\smallline
\hangindent=\parindent
Silk, J., \& White, S.~D.~M.~1978, \apjl, 226, L103
\smallline
\hangindent=\parindent
Stark, A.~A., Heiles, C., Bally, J., \& Linke, R.~1984, privately
distributed magnetic tape
\smallline
\hangindent=\parindent
Tanaka, Y., Inoue, H., \& Holt, S.~S.~1994, \pasj, 46, L37
\smallline
\hangindent=\parindent
Tr\"umper, J.~1983, Adv.~Space Res., 2(4), 241
\smallline
\hangindent=\parindent
van den Bergh, S.~1995, \apjl, 453, L55
\smallline
\hangindent=\parindent
Weinberg, S.~1972, Gravitation and Cosmology (Wiley: New York), 485
\smallline
\hangindent=\parindent
Yamashita, K.~1994, in New Horizon of X-Ray Astronomy---First Results
from \asca, ed.\ F.~Makino \& T.~Ohashi (Tokyo: Universal Academy),
279

\vfill\eject

$$\vbox{\halign to 4.5 truein {\tabskip=0.5em plus1em minus 1em \qquad
# \hfil & \hfil # \hfil & \hfil # \hfil \cr
\multispan3{\hfil {\sc TABLE 1} \hfil} \cr
\noalign{\vskip 5pt}
\multispan3{\hfil {\sc Circular Isothermal-$\beta$ Model Fit} \hfil}\cr
\noalign{\vskip 5pt}
\noalign{\hrule}
\noalign{\vskip 2pt}
\noalign{\hrule}
\noalign{\vskip 5pt}
 Parameter & Fitted Value & Error $^{\rm a}$  \cr
\noalign{\vskip 5pt}
\noalign{\hrule}
\noalign{\vskip 5pt}
\multispan3{\hfil PSPC X-ray Data\hfil}\cr
\noalign{\vskip 2pt}
 R.A.\ (J2000)       & $0^{\rm h} 18^{\rm m} 33.18^{\rm s}$ 
  & $\pm 0.05^{\rm s}$                \cr
 Decl.\ (J2000)       & $16^\circ 26^\prime  17\arcsec9$
  & $\pm 0.8^{\prime\prime}$           \cr
 $\beta$                & 0.728
  & $^{+0.025}_{-0.022}$               \cr
 $\theta_C$ ($^\prime$) & 0.679
  & $^{+0.045}_{-0.039}$               \cr
 $b_{X}(0)$ (\cpspamin)   & $4.71 \times 10^{-2}$
  & $\pm0.24 \times 10^{-2}$  \cr
\noalign{\vskip 5pt}
\multispan3{\hfil OVRO SZ Effect Data\hfil}\cr
\noalign{\vskip 2pt}
 $\Delta T_{RJ}(0)$ ($\mu$K)  &  $-1205$  & $\pm 190$ \cr
\noalign{\vskip 5pt}
\noalign{\hrule}
\noalign{\vskip 5pt}
\multispan3{$^{\rm a}$ Single parameter 1-$\sigma$ errors\hfil}\cr}}$$

\vfill\eject

$$\vbox{\halign to 5.5  truein {\tabskip=0.5em plus1em minus 1em \qquad
# \hfil & \hfil # \hfil & \hfil # \hfil \cr
\multispan3{\hfil {\sc TABLE 2} \hfil} \cr
\noalign{\vskip 5pt}
\multispan3{\hfil {\sc Elliptical Isothermal-$\beta$ Model Fit} \hfil}\cr
\noalign{\vskip 5pt}
\noalign{\hrule}
\noalign{\vskip 2pt}
\noalign{\hrule}
\noalign{\vskip 5pt}
 Parameter & Fitted Value & Error $^{\rm a}$  \cr
\noalign{\vskip 5pt}
\noalign{\hrule}
\noalign{\vskip 5pt}
\multispan3{\hfil PSPC X-ray Data\hfil}\cr
\noalign{\vskip 2pt}
 R.A.\ (J2000)        & $0^{\rm h} 18^{\rm m} 33.18^{\rm s}$ 
  & $\pm 0.05^{\rm s}$                \cr
 Decl.\ (J2000)       & $16^\circ 26^\prime 17\arcsec8$
  & $\pm 0.8^{\prime\prime}$           \cr
 $\beta$              & 0.737
  & $^{+0.027}_{-0.022}$               \cr
 $\theta_C$ ($^\prime$) (along major axis) & 0.746
  & $\pm 0.044$               \cr
 $e$ $^{\rm b}$             & 1.176 & $^{+0.033}_{-0.030}$ \cr
 Position angle ($^\circ$)  & 50.8  & $\pm4.9$             \cr
 $b_{X}(0)$ (\cpspamin) & $4.72 \times 10^{-2}$
  & $\pm0.24 \times 10^{-2}$  \cr
\noalign{\vskip 5pt}
\multispan3{\hfil OVRO SZ Effect Data\hfil}\cr
\noalign{\vskip 2pt}
 $\Delta T_{RJ}(0)$ ($\mu$K)  &  $-1207$  & $\pm 190$ \cr
\noalign{\vskip 5pt}
\noalign{\hrule}
\noalign{\vskip 5pt}
\multispan3{$^{\rm a}$ Single parameter 1-$\sigma$ errors\hfil}\cr
\multispan3{$^{\rm b}$ Defined as observed  major axis divided by minor
axis\hfil}\cr}}$$ 

\vfill\eject

$$\vbox{\halign to 6 truein {\tabskip=0.5em plus1em minus 1em \qquad
# \hfil & \qquad \hfil # \hfil & \hfil # \hfil \cr
\multispan3{\hfil {\sc TABLE 3} \hfil} \cr
\noalign{\vskip 5pt}
\multispan3{\hfil {\sc Joint PSPC/GIS Spectral Model Fit}\hfil}\cr 
\noalign{\vskip 5pt}
\noalign{\hrule}
\noalign{\vskip 2pt}
\noalign{\hrule}
\noalign{\vskip 5pt}
 Parameter & Fitted Value & Error $^{\rm a}$  \cr
\noalign{\vskip 5pt}
\noalign{\hrule}
\noalign{\vskip 5pt}
\multispan3{\hfil Cluster: CL0016+16 (Thermal Model)\hfil}\cr
\noalign{\vskip 2pt}
 $kT$ (keV) $^{\rm b}$         & $7.55$ & $^{+0.72}_{-0.58}$ $^{\rm c}$ \cr
 [Fe]/[H] $^{\rm d}$           & $0.07$ & $^{+0.11}_{-0.07}$ \cr
 $N_{\rm H}$ (atoms cm$^{-2}$) & $5.59 \times 10^{20}$
   & $^{+0.41}_{-0.36} \times 10^{20}$ \cr
 Emission Measure $^{\rm e}$ (cm$^{-5}$) & $3.65 \times 10^{11}$
   & $\pm0.09 \times 10^{11}$ \cr
\noalign{\vskip 5pt}
\multispan3{\hfil AGN: QSO 0015+162 (Power-law Model)\hfil}\cr
\noalign{\vskip 2pt}
 $\alpha_p$                    & $2.54$ & $\pm0.18$ \cr
 $F_{1 \rm keV}$ (photons keV$^{-1}$ cm$^{2}$ s$^{-1}$)  
   & $5.14\times 10^{-5}$ & $^{+0.43}_{-0.39} \times 10^{-5}$ \cr
\noalign{\vskip 5pt}
\noalign{\hrule}
\noalign{\vskip 5pt}
\multispan3{$^{\rm a}$ Single parameter 1-$\sigma$ errors\hfil}\cr
\multispan3{$^{\rm b}$ Source frame for $z=0.5455$\hfil}\cr
\multispan3{$^{\rm c}$ RSS of $^{+0.63}_{-0.48}$ (statistical error) 
and $^{+0.35}_{-0.33}$ ($\pm$6\% background subtraction error) \hfil}\cr
\multispan3{$^{\rm d}$ Relative to solar [Fe]/[H] $= 4\times 10^{-5}$ \hfil}\cr
\multispan3{$^{\rm e}$ Flux (0.2--2.0 keV) $= 1 \times 10^{-12}$ ergs
cm$^{-2}$ s$^{-1}$ \hfil}\cr
\multispan3{$\phantom{^{\rm d}}$ Flux (2--10 keV) $= 1.5 \times 10^{-12}$ ergs
cm$^{-2}$ s$^{-1}$ \hfil}\cr
}}$$

\vfill\eject

$$\vbox{\halign to 6.5truein { \tabskip=0.5em plus1em minus 1em
  \quad # \hfil
  & \hfil # \hfil
  & \hfil # \hfil 
  & \hfil # \hfil 
  & \hfil # \hfil
  & \hfil # \hfil 
  & \hfil # \hfil 
  & \hfil # \hfil \cr  
\multispan8{\hfil {\sc TABLE 4} \hfil} \cr
\noalign{\vskip 5pt}
\multispan8{\hfil {\sc Uncertainty in $H_0$ from Observational Errors} \hfil}\cr
\noalign{\vskip 5pt}
\noalign{\hrule}
\noalign{\vskip 2pt}
\noalign{\hrule}
\noalign{\vskip 5pt}
      &  $kT$ & [Fe]/[H]          & $N_{\rm H}$ 
  & $\beta$ & $\theta_C$ & $b_{X}(0)$ & $\Delta T_{RJ}(0)$ \cr
\noalign{\vskip 5pt}
\noalign{\hrule}
\noalign{\vskip 5pt}
Observational Error $^{\rm a}$ (\%) & $^{+9.5}_{-7.7}$ & $^{+157}_{-100}$ 
  & $^{+7.3}_{-6.4}$ & $^{+3.4}_{-3.0}$ & $^{+6.6}_{-5.7}$ 
  & $\pm 5.1$ & $\pm 15.7$ \cr
\noalign{\vskip 5pt}
 $\delta H_0$  $^{\rm a}$ \kmsMpc
   & $^{+9.5}_{-7.0}$ & $^{-1.0}_{+0.6}$ & $\pm0.5$ 
   & \multispan3{\leaders\hrule height2.4pt depth-2.0pt\hfill 
     $\,\,\pm 2.0$ $^{\rm b}\!$ 
     \leaders\hrule height2.4pt depth-2.0pt\hfill} & 
 $^{-11.8}_{+18.8}$ \cr 
\noalign{\vskip 5pt}
\noalign{\hrule}
\noalign{\vskip 5pt}
\multispan8{$^{\rm a}$ 1-$\sigma$ errors  \hfil}\cr
\multispan8{$^{\rm b}$ Over three-dimensional error surface for
$\beta$, $\theta_C$, and $b_{X}(0)$ \hfil}\cr
}}$$

\vskip 10 pt

$$\vbox{\halign to 4.5truein {\tabskip=0.5em plus1em minus 1em
  \qquad # \hfil
  & \hfil # \hfil 
  & \hfil # \hfil \cr  
\multispan3{\hfil {\sc TABLE 5} \hfil} \cr
\noalign{\vskip 5pt}
\multispan3{\hfil {\sc Uncertainty in $H_0$ from Unknown Geometry} \hfil}\cr
\noalign{\vskip 5pt}
\noalign{\hrule}
\noalign{\vskip 2pt}
\noalign{\hrule}
\noalign{\vskip 5pt}
                             & \multispan2{\hfil $H_0$ (\kmsMpc)\hfil} \cr
\noalign{\vskip -7pt}
                             & \multispan2{\hrulefill} \cr
\noalign{\vskip -2pt}
Model                        & Oblate & Prolate \cr
\noalign{\vskip 5pt}
\noalign{\hrule}
\noalign{\vskip 5pt}
Circular                     &  \multispan2{\hfil 47.2\hfil} \cr
Inclination angle $90^\circ$  &  50.7 & 43.1 \cr
Inclination angle $70^\circ$  &  49.4 & 43.9 \cr
Inclination angle $50^\circ$  &  43.3 & 47.2 \cr
Intrinsic major/minor axis ratio  1.5    &  39.7 & 55.0 \cr
\noalign{\vskip 5pt}
\noalign{\hrule}
\noalign{\vskip 5pt}
}}$$

\vfill\eject

$$\vbox{\halign to 6 truein {\tabskip=0.5em plus1em minus 1em \qquad
# \hfil &  \qquad \hfil # \hfil & \hfil # \hfil & # \hfil \cr
\multispan4{\hfil {\sc TABLE 6} \hfil} \cr
\noalign{\vskip 5pt}
\multispan4{\hfil {\sc Summary of X-Ray/SZ Effect $H_0$ Measurements for}\hfil}\cr 
\multispan4{\hfil {\sc Spherically Symmetric, Isothermal Cluster Models}\hfil}\cr 
\noalign{\vskip 5pt}
\noalign{\hrule}
\noalign{\vskip 2pt}
\noalign{\hrule}
\noalign{\vskip 5pt}
             &        &  $H_0$ $^{\rm a}$     &  \cr
Cluster      &   $z$  &  (\kmsMpc) & Reference \cr
\noalign{\vskip 5pt}
\noalign{\hrule}
\noalign{\vskip 5pt}
Coma & 0.0232 & $64^{+25}_{-21}$ $^{\rm b}$  & Myers et al.~1997 \cr
Abell 2256   & 0.0581 & $69^{+21}_{-18}$  & Myers et al.~1997 \cr
Abell 478    & 0.0881 & $31^{+17}_{-13}$  & Myers et al.~1997 \cr
Abell 2142   & 0.0899 & $46^{+41}_{-28}$  & Myers et al.~1997 \cr
Abell 2218   & 0.171  & $59 \pm 23$       & Birkinshaw \& Hughes 1994 \cr
Abell 2218   & 0.171  & $34^{+16}_{-14}$  & Jones 1995 \cr
Abell 665    & 0.182  & $46 \pm 16$       & Hughes \& Birkinshaw 1998 \cr
Abell 2163   & 0.201  & $58^{+39}_{-22}$  & Holzapfel et al.~1997 \cr
\cl\         & 0.5455 & $47^{+23}_{-15}$  & This work \cr
\noalign{\vskip 5pt}
Ensemble average $^{\rm c}$ & $\ldots$ & $47.1 \pm 6.8$  & $\ldots$ \cr
\noalign{\vskip 5pt}
\noalign{\hrule}
\noalign{\vskip 5pt}
\multispan4{$^{\rm a}$ Assuming $\Omega_0=0.2,\lambda_0=0$;
uncertainties quoted at 1 $\sigma$  \hfil}\cr
\multispan4{$^{\rm b}$ Derived assuming a nonisothermal cluster
temperature distribution  \hfil}\cr 
\multispan4{$^{\rm c}$ Not including Coma  \hfil}\cr
}}$$

\vfill\eject

{\bf\centerline{FIGURE CAPTIONS}}
\medline
\item{Fig.~1---}A portion of the 0.4-2.4 keV \rosat\ PSPC X-ray image
centered on the distant cluster \cl\ ($z=0.5455$). Coordinates are
quoted in epoch J2000. The grayscale 
shows the number of raw counts detected in each $7\arcsec5$
square pixel ranging from a minimum of 1 (in source-free regions of
the map) to a peak of 38 (near the center of the cluster). The
contours show the background subtracted, exposure corrected data after
adaptive smoothing.  Contour levels start at $2.4 \times 10^{-4}$
counts s$^{-1}$ arcmin$^{-2}$ (approximately the average background
level near the center of the detector) and increase by multiplicative
factors of 1.75.   The bright X-ray source immediately to the north of
the cluster is an AGN, QSO 0015+162, at a redshift $z = 0.554$
(Margon et al.~1983). The recently discovered poor
cluster, \rx, that is a companion to \cl\ (Hughes et al.~1995), is
partially visible to the southwest.
\medline
\item{Fig.~2---}{\it ROSAT} PSPC and {\it ASCA} GIS spectral data and
best-fit models.  The cluster was fit with a thermal emission model
with variable abundances and line-of-sight absorption.  The AGN, QSO
0015+162, was assumed to have a power-law spectral form with the same
absorption as the cluster. \cl\ and the AGN were resolved in the {\it
ROSAT} data, so separate PSPC spectra were extracted and fit.
However, these sources were not resolved by {\it ASCA} and only a
single GIS spectrum was extracted (from a 5$^\prime$ radius region
centered on \cl).  The sum of the cluster model and AGN model was fit
to the GIS data.
\medline
\item{Fig.~3---}The \SZ\ effect data for several points on a NS line
through \cl. The declination offsets for each point are measured
relative to a nominal cluster center at 00$^{\rm h}$18$^{\rm m}$34$^{\rm
s}$, 16$^\circ$26$^\prime$20$^{\prime\prime}$. The error bars contain
contributions from the uncertainties in point-by-point systematic
errors, such as the radio source corrections that have been applied
(and which are large for the southernmost two points in the scan). The
horizontal lines indicate the $\pm 1\sigma$ error range on the
estimated zero level offset. The best-fit spherical model for the
cluster gas consistent with the X-ray image is shown by the dashed
lines, which correspond to the best-fit normalization $N_{\rm RJ}$ and
its $\pm 1\sigma$ errors.
\medline
\item{Fig.~4---}Comparison of the radial X-ray profiles of the
best-fitting elliptical isothermal-$\beta$ model with the
\rosat\ PSPC data integrated within azimuthal quadrants as labelled.
The background level is indicated by the horizontal dashed line, the
cluster emission is the dotted curve, and the solid histogram is the
sum of these. The cluster emission is above background within a radius
of $\sim$3$^\prime$.  The smaller panels beneath each radial surface
brightness plot show the residuals (in units of $\sigma$: data minus
model divided by the statistical error) from the fits of
the elliptical isothermal-$\beta$ model.
\medline
\item{Fig.~5---}The effect of the cosmological parameters $\Omega_0$
and $\lambda_0$ on the derived value of $H_0$.  The function $f$
parameterizes the ratio of $H_0$ determined assuming a Friedmann model
with $\Omega_0 = q_0 = 0$ to that with arbitrary $\Omega_0$ and
cosmological constant $\lambda_0$.
\medline
\item{Fig.~6---}Variation of the derived value of the Hubble constant,
$H_0$, with the intrinsic major/minor axis ratio of \cl\ for oblate
and prolate geometries. This plot assumes that the three dimensional
structure of the cluster is given by an ellipsoid of revolution with
symmetry axis lying at different angles of inclination to the line of
sight.  For intrinsic major/minor axis ratio values on the left hand
side of the figure, the symmetry axis lies near the plane of the sky
($i=90^\circ$). For an intrinsic axis ratio of 1.5 (indicated by the
vertical dotted line) the inclination angles are $i=45^\circ$ (oblate)
and $i=34^\circ$ (prolate).

\vfill\eject

\centerline{\epsfbox{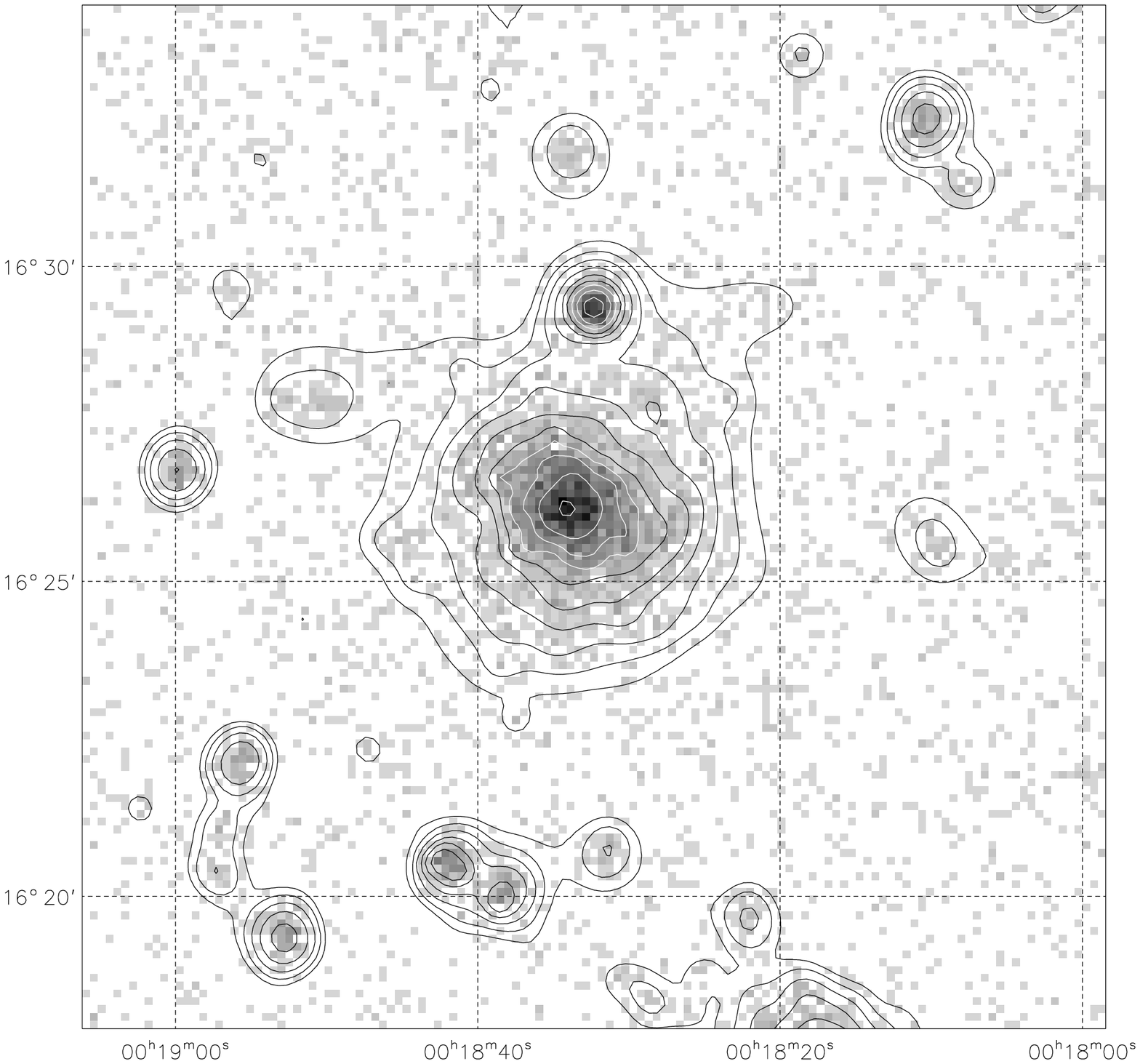}}

\vskip 20pt
\line{\hfill Figure 1}
\vfill\eject

\centerline{\epsfbox{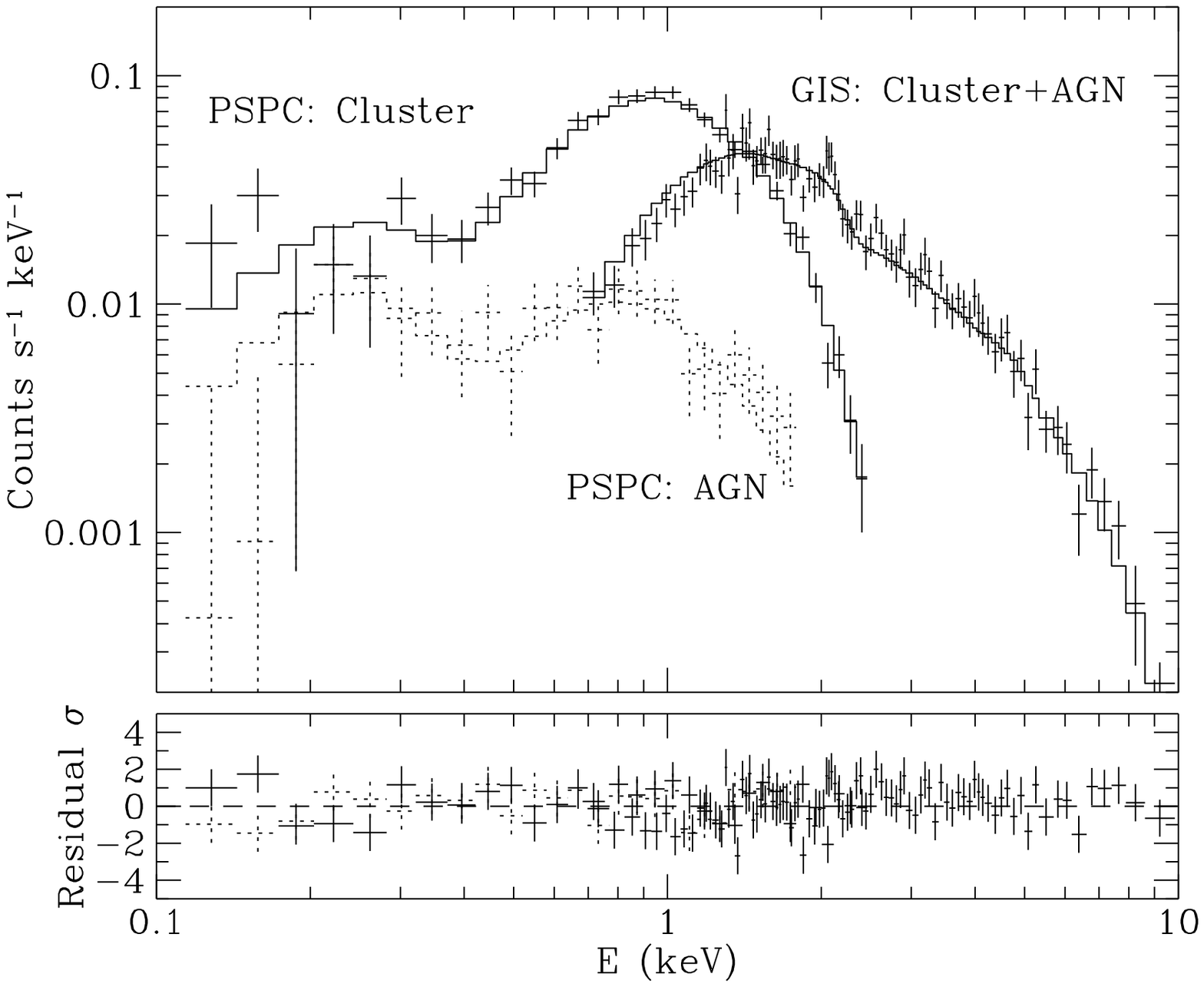}}

\vskip 20pt
\line{\hfill Figure 2}

\vfill\eject

\epsfclipon
\centerline{\epsfbox{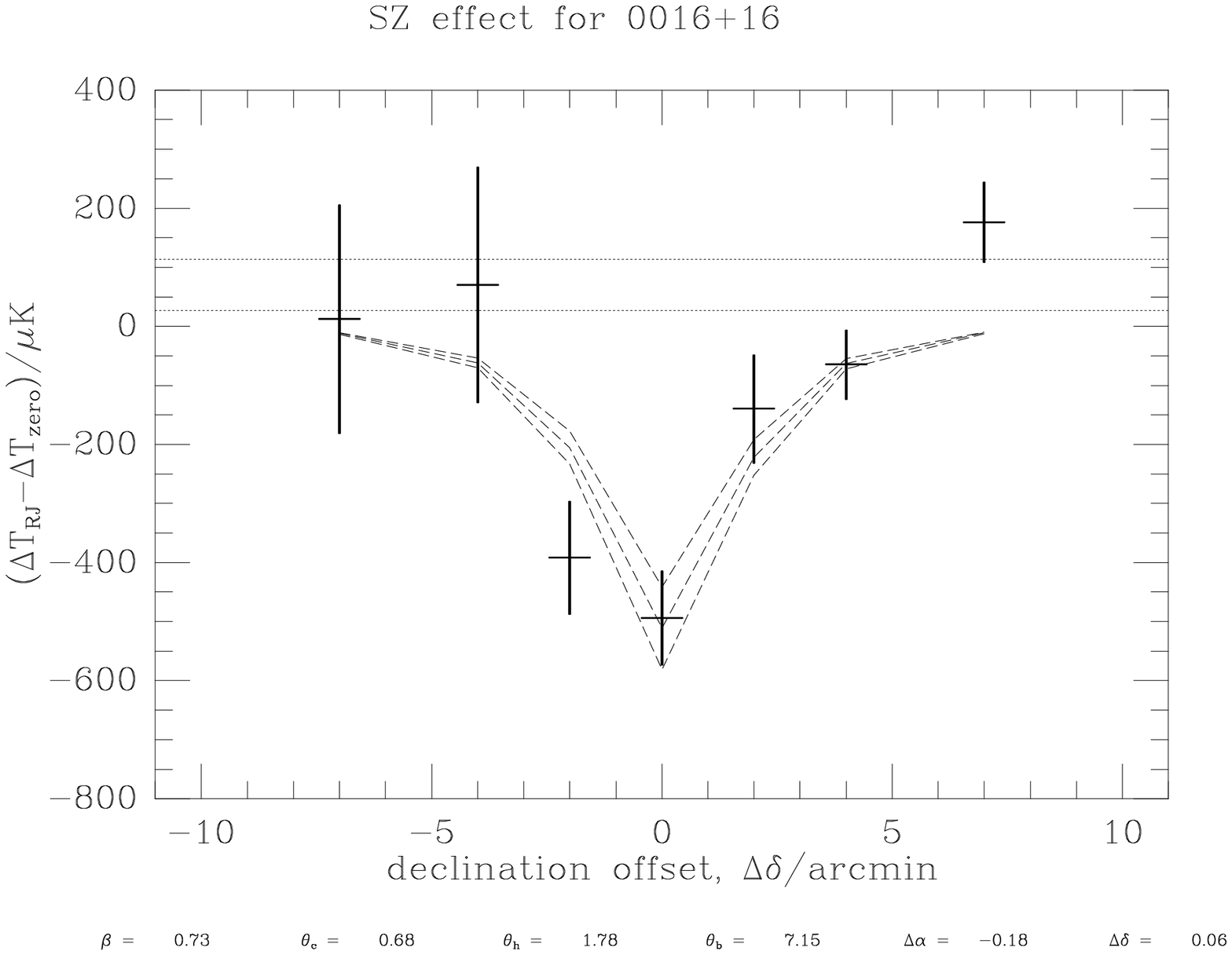}}
\epsfclipoff

\vskip 20pt
\line{\hfill Figure 3}

\vfill\eject

\centerline{\epsfbox{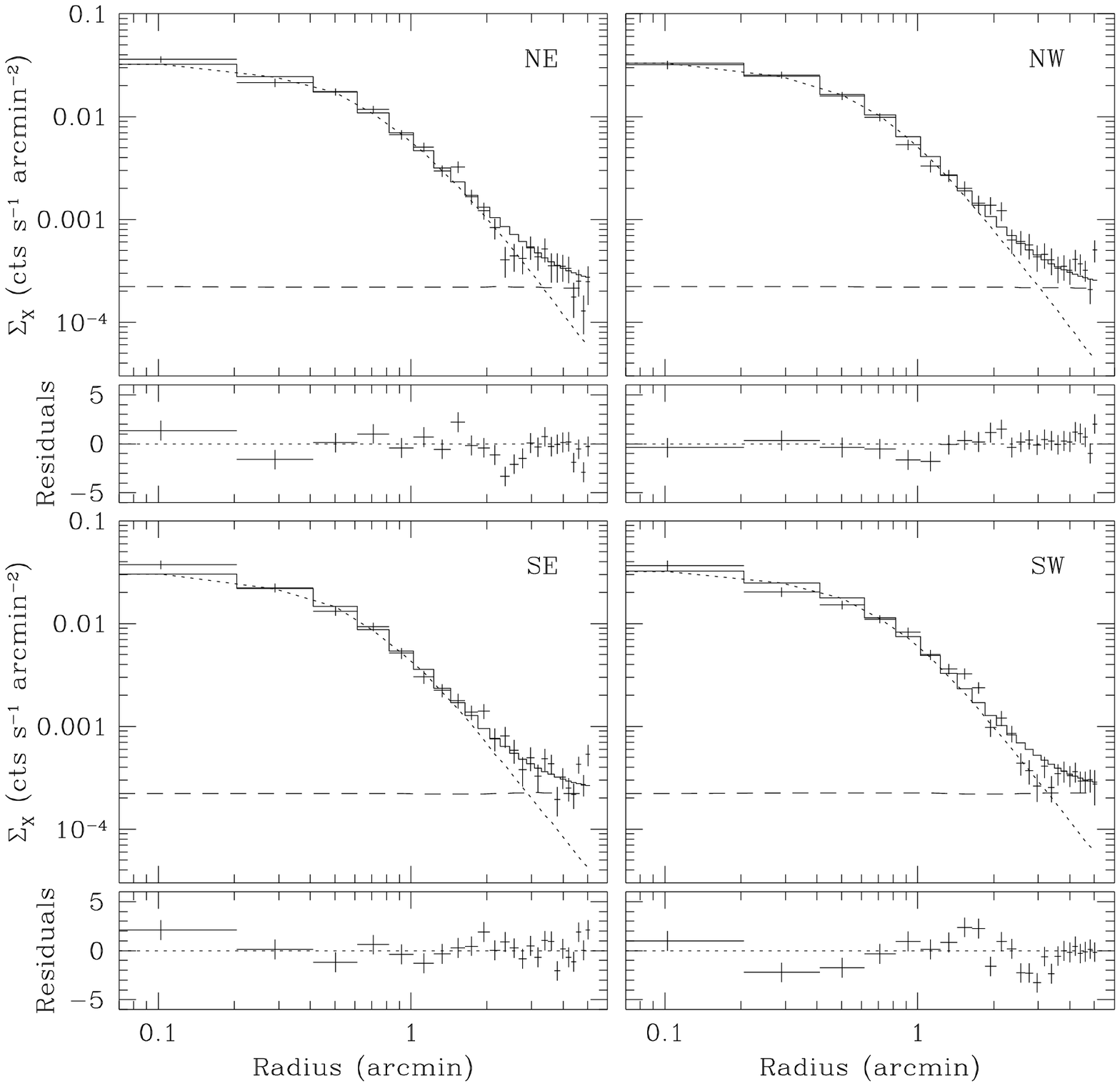}}

\vskip 20pt
\line{\hfill Figure 4}

\vfill\eject

\centerline{\epsfbox{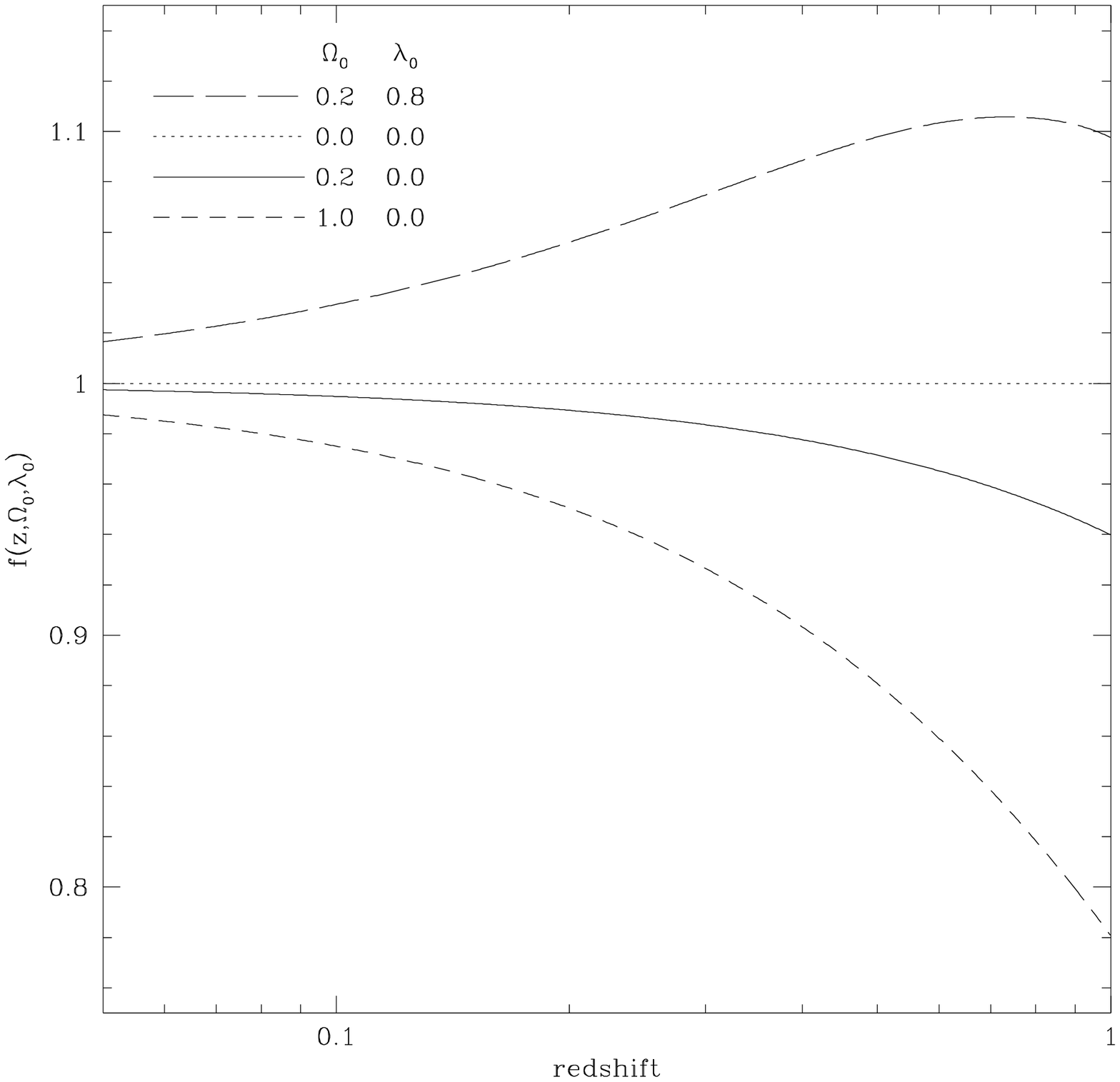}}

\vskip 20pt
\line{\hfill Figure 5}

\vfill\eject

\centerline{\epsfbox{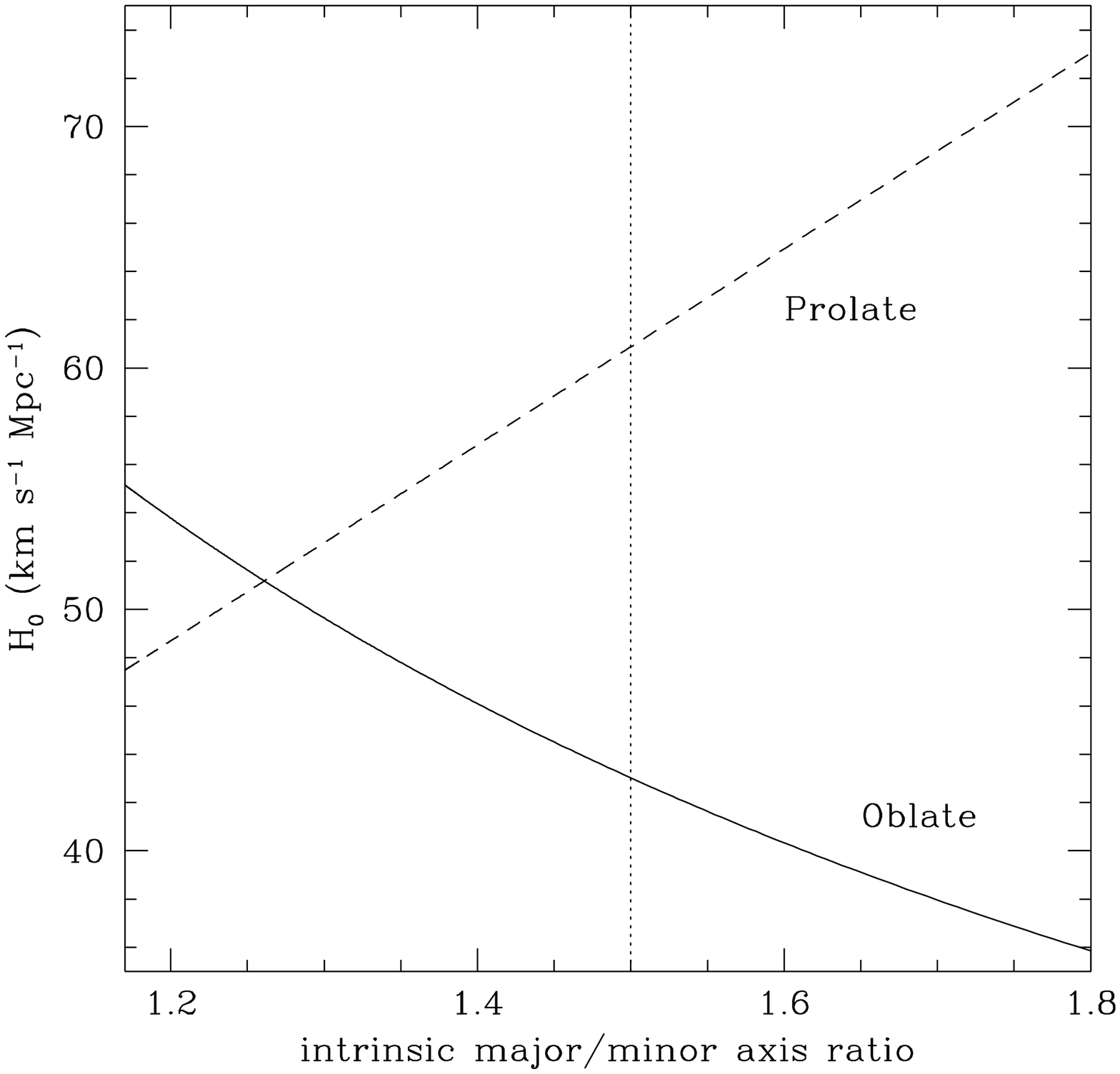}}

\vskip 20pt
\line{\hfill Figure 6}

\vfill\eject\end